\documentclass[twocolumn,twocolappendix,times]{aastex63}

\usepackage{CJK}    
\usepackage{multirow}   
\usepackage{amsmath}
\usepackage{hyperref}
\hypersetup{
    colorlinks=true,
    linkcolor=blue,
    filecolor=blue,      
    urlcolor=blue,
}
\newcommand{\herschel}{{\it Herschel\/}}
\newcommand{\chandra}{{\it Chandra\/}}

\newcommand{\xmm}{\hbox{\it XMM-Newton\/}}

\newcommand{\athena}{{\it Athena\/}}
\newcommand{\lynx}{{\it Lynx\/}}
\newcommand{\jwst}{{\it JWST\/}}
\newcommand{\spitzer}{{\it Spitzer\/}}

\newcommand{\xray}{\hbox{X-ray}}  

\newcommand{\lx}{L_{\rm X}}
\newcommand{\lir}{L_{\rm IR}}
\newcommand{\luvo}{L_{\rm UV/O}}
\newcommand{\lhmxb}{L^{\rm HMXB}_{\rm X}}
\newcommand{\llmxb}{L^{\rm LMXB}_{\rm X}}
\newcommand{\dethmxb}{\delta_{\rm HMXB}}
\newcommand{\detlmxb}{\delta_{\rm LMXB}}

\newcommand{\ld}{L_{\rm disk}}

\newcommand{\mstar}{M_{\star}}
\newcommand{\fracA}{{\rm frac_{AGN}}}

\newcommand{\ox}{\alpha_{\rm ox}}
\newcommand{\ang}{\textup{\AA}}

\newcommand{\luvr}{L_{\nu, 2500 \ang}}
\newcommand{\lxr}{L_{\rm \nu, 2 keV}}
\newcommand{\lrr}{L_{\rm \nu, 5 GHz}}
\newcommand{\pagn}{P_{\rm AGN,1.4GHz}}

\newcommand{\xcig}{\hbox{\sc x-cigale}}
\newcommand{\cig}{{\sc cigale}}
\newcommand{\cigv}{\hbox{\sc cigale v2022.0}}

\newcommand{\fst}[1]{{#1}}
\newcommand{\scd}[1]{{#1}}

\shorttitle{CIGALE v2022.0}
\shortauthors{Yang et al.}

\begin{document}
\begin{CJK*}{UTF8}{gbsn}
\title{Fitting AGN/galaxy X-ray-to-radio SEDs with CIGALE and improvement of the code}

\correspondingauthor{G. Yang}
\email{gyang206265@gmail.com}

\author[0000-0001-8835-7722]{Guang Yang (杨光)}
\affiliation{Department of Physics and Astronomy, Texas A\&M
  University, College Station, TX, 77843-4242 USA}
\affiliation{George P.\ and Cynthia Woods Mitchell Institute for
 Fundamental Physics and Astronomy, Texas A\&M University, College Station, TX, 77843-4242 USA}
 
\author{M\'ed\'eric Boquien}
\affiliation{Centro de Astronom\'ia (CITEVA), Universidad de Antofagasta, Avenida Angamos 601, Antofagasta, Chile}

\author{W. N. Brandt}
\affiliation{Department of Astronomy and Astrophysics, 525 Davey Lab, The Pennsylvania State University, University Park, PA 16802, USA}
\affiliation{Institute for Gravitation and the Cosmos, The Pennsylvania State University, University Park, PA 16802, USA}
\affiliation{Department of Physics, 104 Davey Laboratory, The Pennsylvania State University, University Park, PA 16802, USA}

\author{V\'eronique Buat}
\affiliation{Aix Marseille Univ, CNRS, CNES, LAM, Marseille, France}
\affiliation{Institut Universitaire de France (IUF)}

\author{Denis Burgarella}
\affiliation{Aix Marseille Univ, CNRS, CNES, LAM, Marseille, France}

\author{Laure Ciesla}
\affiliation{Aix Marseille Univ, CNRS, CNES, LAM, Marseille, France}

\author{Bret D. Lehmer}
\affiliation{Department of Physics, University of Arkansas, 226 Physics Building, 825 West Dickson Street, Fayetteville, AR 72701, USA}

\author[0000-0003-3080-9778]{Katarzyna Ma{\l}ek}
\affiliation{National Centre for Nuclear Research, ul. Pasteura 7, 02-093 Warszawa, Poland}
\affiliation{Aix Marseille Univ, CNRS, CNES, LAM, Marseille, France}

\author{George Mountrichas}
\affiliation{Instituto de Fisica de Cantabria (CSIC-Universidad de 
Cantabria), Avenida de los Castros, 39005 Santander, Spain}

\author[0000-0001-7503-8482]{Casey Papovich}
\affiliation{Department of Physics and Astronomy, Texas A\&M University, College
Station, TX, 77843-4242 USA}
\affiliation{George P.\ and Cynthia Woods Mitchell Institute for
Fundamental Physics and Astronomy, Texas A\&M University, College
Station, TX, 77843-4242 USA}

\author{Estelle Pons}
\affiliation{Aix Marseille Univ, CNRS, CNES, LAM, Marseille, France}

\author{Marko Stalevski}
\affiliation{Astronomical Observatory, Volgina 7, 11060 Belgrade, Serbia}
\affiliation{Sterrenkundig Observatorium, Universiteit Gent, Krijgslaan 281-S9, Gent, 9000, Belgium}

\author{Patrice Theul\'e}
\affiliation{Aix Marseille Univ, CNRS, CNES, LAM, Marseille, France}

\author{Shifu Zhu}
\affiliation{Department of Astronomy and Astrophysics, 525 Davey Lab, The Pennsylvania State University, University Park, PA 16802, USA}
\affiliation{Institute for Gravitation and the Cosmos, The Pennsylvania State University, University Park, PA 16802, USA}



\begin{abstract}
Modern and future surveys effectively provide a panchromatic view for large numbers of extragalactic objects. 
Consistently modeling these multiwavelength survey data is a critical but challenging task for extragalactic studies.  
The Code Investigating GALaxy Emission (\cig) is an efficient {\sc python} code for spectral energy distribution (SED) fitting of galaxies and active galactic nuclei (AGNs). 
Recently, a major extension of \cig\ (named \xcig) has been developed to account for AGN/galaxy \xray\ emission and improve AGN modeling at UV-to-IR wavelengths.
Here, we apply \xcig\ to different samples, including COSMOS spectroscopic type~2 AGNs, CDF-S \xray\ detected normal galaxies, SDSS quasars, and COSMOS radio objects.
From these tests, we identify several weaknesses of \xcig\ and improve the code accordingly. 
These improvements are mainly related to AGN intrinsic \xray\ anisotropy, \xray\ binary emission, AGN accretion-disk SED shape, and AGN radio emission.
These updates improve the fit quality and allow new interpretation of the results, based on which we discuss physical implications. 
For example, we find that AGN intrinsic \xray\ anisotropy is moderate, and can be modeled as $L_X(\theta) \propto 1+\cos \theta$, where $\theta$ is the viewing angle measured from the AGN axis. 
We merge the new code into the major branch of \cig, and publicly release this new version as \cigv\ on \url{https://cigale.lam.fr}
\vspace{1.5 cm}
\end{abstract}

\section{Introduction}
\label{sec:intro}
Extragalactic surveys from \xray\ to radio have become increasingly important for studying the evolution of galaxies and supermassive black holes (BHs) across cosmic history. 
Broad wavelength coverage provides insights into a diversity of properties of extragalactic sources.
X-rays can reveal intrinsic active galactic nucleus (AGN) emission, even when it is obscured.  
UV/optical light traces young stars and unobscured AGN accretion disks. 
IR light reveals the dust-obscured AGN and/or star-formation (SF) activities.    
Radio emission can be generated by high-energy electrons associated with, e.g., AGN jets, AGN-driven winds, and HII regions. 

Modern surveys such as LSST \citep{ivezic19} and eRASS \citep{predehl21} can sample millions-to-billions of diverse objects, from luminous quasars to low-luminosity AGNs, and from brightest cluster galaxies (BCGs) to dwarf galaxies.
Interpreting these large volumes of multiwavelength data coherently and efficiently is a challenging task for extragalactic studies.   

Many codes have been developed to fit AGN/galaxy spectral energy distributions 
(SEDs; see Fig.~1 of \citealt{thorne20} for a summary of different codes). 
The Code Investigating GALaxy Emission (\cig) is an open-source SED-fitting code written in {\sc python} \citep{burgarella05, boquien19}. 
It employs a parallel algorithm, able to build thousands of SED models per second and fit them to data.
The SED models are built through a series of ``modules'' defined by the user. 
This architecture is designed to allow easy updates or even the addition of branches in the code that carries scientific investigations.
For example, the dust-attenuation module applies a specific attenuation recipe to the starlight and line emission, and the dust-emission module is responsible for the IR dust radiation. 
The dust-emission module always normalizes the SED so that the re-emitted total energy is equal to the obscured total energy in the dust-attenuation module. 
In this way, \cig\ obeys the law of energy conservation. 
\cig\ has an AGN module that is responsible for the UV-to-IR emission from AGNs \citep{boquien19, ciesla15}.  

Recently, \cite{yang20} developed a major \cig\ extension, \xcig, adding a brand-new range of the electromagnetic spectrum (i.e., X-ray) to the existing UV-to-radio range.
\cite{yang20} also implemented several AGN-related improvements including a clumpy torus model and a polar-dust model for \xcig. 
The \xray\ module allows the modeling of \xray\ fluxes, accounting for the emission 
from both AGNs and galaxies (i.e., hot gas and \xray\ binaries). 
\xcig\ has become increasingly popular especially among AGN researchers \citep[e.g.,][]{zou20,mountrichas21,ni21,toba21,yang21}.

In this work, we aim at testing \xcig\ on diverse extragalactic populations, therefore we use several AGN/galaxy samples selected over different wavelength ranges, including COSMOS spectroscopic type~2 AGNs, CDF-S (\chandra\ Deep Field-South) \xray\ detected normal galaxies, Sloan Digital Sky Survey (SDSS) quasars, and COSMOS radio objects. 
From these tests, we identify weaknesses and improve the code accordingly. 
These improvements are mainly related to AGN \xray\ anisotropy, binary \xray\ emission, AGN accretion-disk SED shape, and AGN radio emission.
We discuss the physical implications based on the fitting results of the new code.   
Finally, we merge the new code into the main branch of \cig, after minimizing the differences between 
the two branches in terms of, e.g., coding structures and variable naming. 
This procedure removes a heavy burden of software maintenance, because, previously, 
an upgrade (such as algorithm improvements and additional functionalities) in 
\cig\ had to be modified and tested before implementation into \xcig, and vice versa. 
We publicly release the merged software as \cigv\ on the \cig\ official website, \url{https://cigale.lam.fr}

The structure of this paper is as follows.
In \S\ref{sec:xray_ani}, we fit a sample of type~2 AGNs and implement AGN anisotropic \xray\ emission. 
In \S\ref{sec:xrb}, we fit a sample of \xray\ detected normal galaxies (non-AGNs) and introduce a flexible recipe for binary \xray\ emission. 
In \S\ref{sec:flex_disk}, we fit a sample of type~1 quasars and implement code changes allowing for more flexible AGN disk SED shapes. 
In \S\ref{sec:agn_radio}, we fit a sample of radio sources and introduce an AGN radio component to the radio module. 
In \S\ref{sec:misc}, we present some miscellaneous updates of the code.  
We summarize our results and discuss future prospects in \S\ref{sec:sum}.

Throughout this paper, we assume a cosmology with $H_0=70$~km~s$^{-1}$~Mpc$^{-1}$, $\Omega_M=0.3$, and $\Omega_{\Lambda}=0.7$.
We adopt a Chabrier initial mass function (IMF; \citealt{chabrier03}).
Quoted uncertainties are at the $1\sigma$\ (68\%) confidence level.
Quoted optical/infrared magnitudes are AB magnitudes.
We adopt the ``Bayesian-like'' (rather than the best-fit) quantities in {\sc (x)-cigale} output catalogs, unless otherwise stated. 
A Bayesian-like quantity/error are calculated as the average/standard deviation of all model values weighted by the probability distribution \citep{noll09, boquien19}. 

\section{AGN \xray\ Anisotropy}
\label{sec:xray_ani}
\subsection{Motivation}
\label{sec:mot_xray_ani}
It is generally believed that the observed \hbox{X-rays} are a result of 
a ``disk-corona'' structure.
The disk emits UV/optical photons, and a fraction of them are up-scattered
to \xray\ wavelengths by the high-energy electrons in the corona 
(i.e., inverse Compton scattering).
The angular dependence of the \xray\ emission is related to the detailed physical properties of the corona, such as shape and optical depth \citep[e.g.,][]{sunyaev85, xu15b}.  

The observations of \cite{liu14} found that type~2 AGNs tend to systematically have lower intrinsic \xray\ luminosity ($\lx$) than type~1 AGNs at a given {\sc [o iv]}~25.89~$\mu$m luminosity. 
Assuming that the {\sc [o iv]} emission from the narrow-line region (NLR) is isotropic, they interpreted their result as an indicator of AGN \xray\ anisotropy, because type~2 AGNs have larger viewing angles (as measured from the AGN axis) than type~1 AGNs under the scheme of AGN unification \citep[e.g.,][]{antonucci93, urry95, netzer15}.
Also, based on the \xray\ and high-resolution mid-IR observations of nearby AGNs, \cite{asmus15} suggested that AGN \xray\ emission might be anisotropic (see their section 5.4 for details). 

However, {\sc x-cigale} assumes that AGN intrinsic \xray\ emission is isotropic, and does not allow anisotropic modeling. 
This assumption could overestimate the \xray\ emission for type~2 viewing angles. 


\subsection{Sample and preliminary fitting}
\label{sec:ani_sample}

\begin{table*}
\centering
\caption{Model parameters for the type~2 AGNs in COSMOS}
\label{tab:par_type2}
\begin{tabular}{llll} \hline\hline
Module & Parameter & Symbol & Values \\
\hline
\multirow{2}{*}{\shortstack[l]{Star formation history\\
                               $\mathrm{SFR}\propto t \exp(-t/\tau)$ }}
    & Stellar e-folding time & $\tau_{\rm star}$ & 0.1, 0.5, 1, 5 Gyr\\
    & Stellar age & $t_{\rm star}$  
            & 0.5, 1, 3, 5, 7 Gyr\\ 
\hline
\multirow{2}{*}{\shortstack[l]{Simple stellar population\\ 
    \cite{bruzual03}}}
    & Initial mass function & $-$ & \cite{chabrier03} \\
    & Metallicity & $Z$ & 0.02 \\
\hline
\multirow{2}{*}{\shortstack[l]{Dust attenuation \\ 
                \cite{calzetti00} }}
    & \multirow{2}{*}{Color excess} & 
        \multirow{2}{*}{$E(B-V)$} &
        \multirow{2}{*}{\shortstack[l]{0.05, 0.1, 0.2, 0.3, 0.4, \\ 
                                       0.5, 0.7, 0.9 mag}} \\\\
\hline
\multirow{2}{*}{\shortstack[l]{Galactic dust emission: \\ \cite{dale14}}}
    & \multirow{2}{*}{\shortstack[l]{Slope in $dM_{\rm dust} \propto U^{-\alpha} dU$}}
    & \multirow{2}{*}{\shortstack[l]{$\alpha$}}
    & \multirow{2}{*}{\shortstack[l]{2}}
    \\\\
\hline
\multirow{3}{*}{\shortstack[l]{AGN (UV-to-IR) \\ SKIRTOR }}
    & AGN contribution to IR luminosity & $\fracA$ & 0--0.99 (step 0.1)  \\
    & Viewing angle & $\theta$ & 60, 70, 80, 90$^\circ$ \\
    & \multirow{1}{*}{\shortstack[l]{Polar-dust color excess}} & \multirow{1}{*}{\shortstack[l]{$E(B-V)_{\rm PD}$}} & \multirow{1}{*}{\shortstack[l]{0, 0.2, 0.4}} \\
\hline
    \multirow{3}{*}{\shortstack[l]{X-ray}}
    & AGN photon index & $\Gamma$ & 1.8 \\
    & Maximum deviation from the $\ox$-$\luvr$ relation & $|\Delta \ox|_{\rm max}$ & 0.2 \\
    & \textbf{AGN \xray\ angle coefficients} & $\boldsymbol{(a_1, a_2)}$ & \textbf{(0, 0) / (0.5, 0) / (1, 0) / (0.33, 0.67)}$^a$ \\
\hline
\end{tabular}
\begin{flushleft}
{\sc Note.} --- For parameters not listed here, we use the default values.
\textbf{Bold font} indicates new parameters in \cigv\ introduced in this work.
(a) Each set of angle coefficients is for one \xcig\ run. 
$(a_1, a_2)=(0,0)$ indicates the \xcig\ (isotropic) run. 
\end{flushleft}
\end{table*}

\begin{figure*}
    \centering
	\includegraphics[width=\columnwidth]{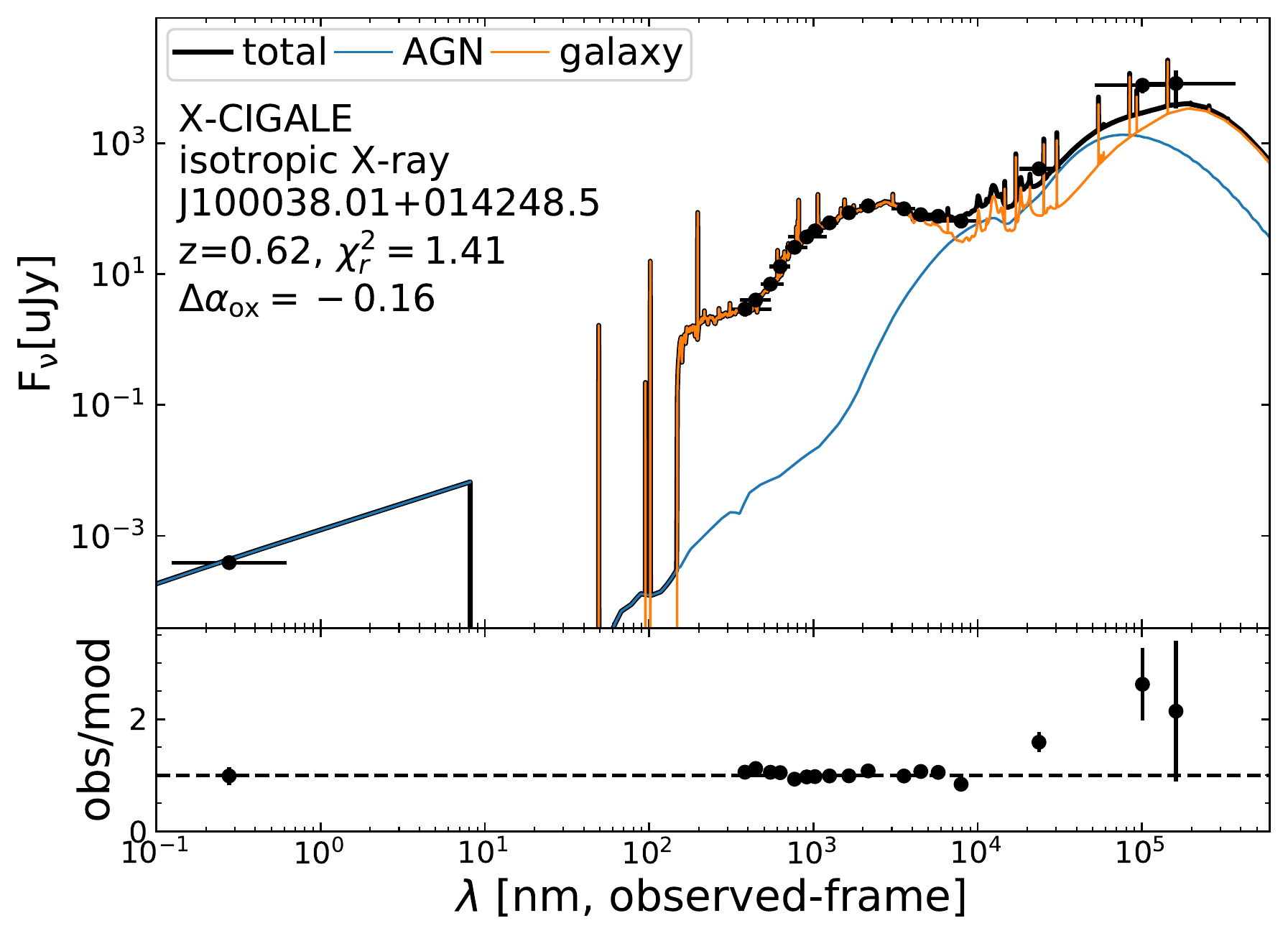}
	\includegraphics[width=\columnwidth]{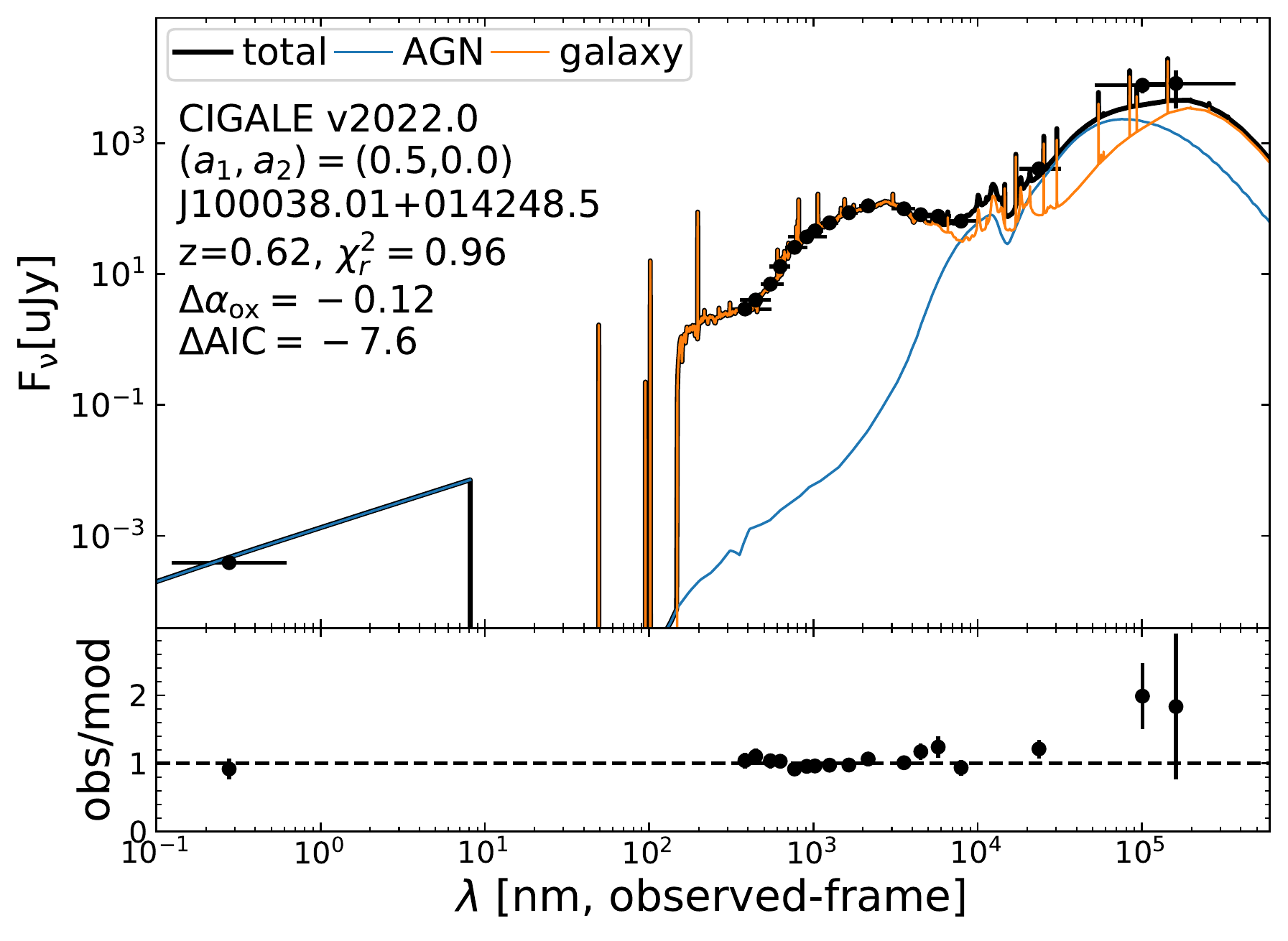}
	\includegraphics[width=\columnwidth]{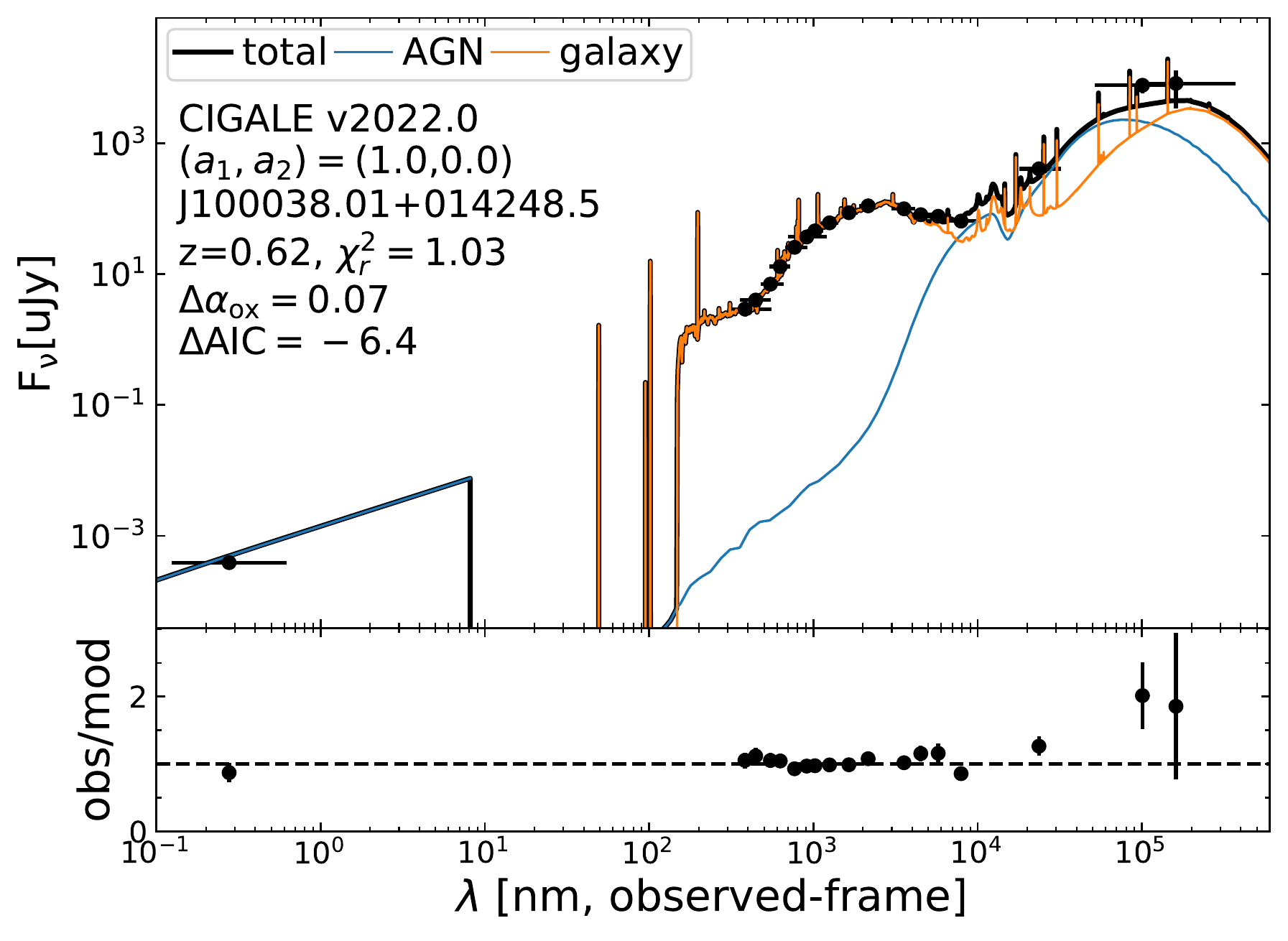}
	\includegraphics[width=\columnwidth]{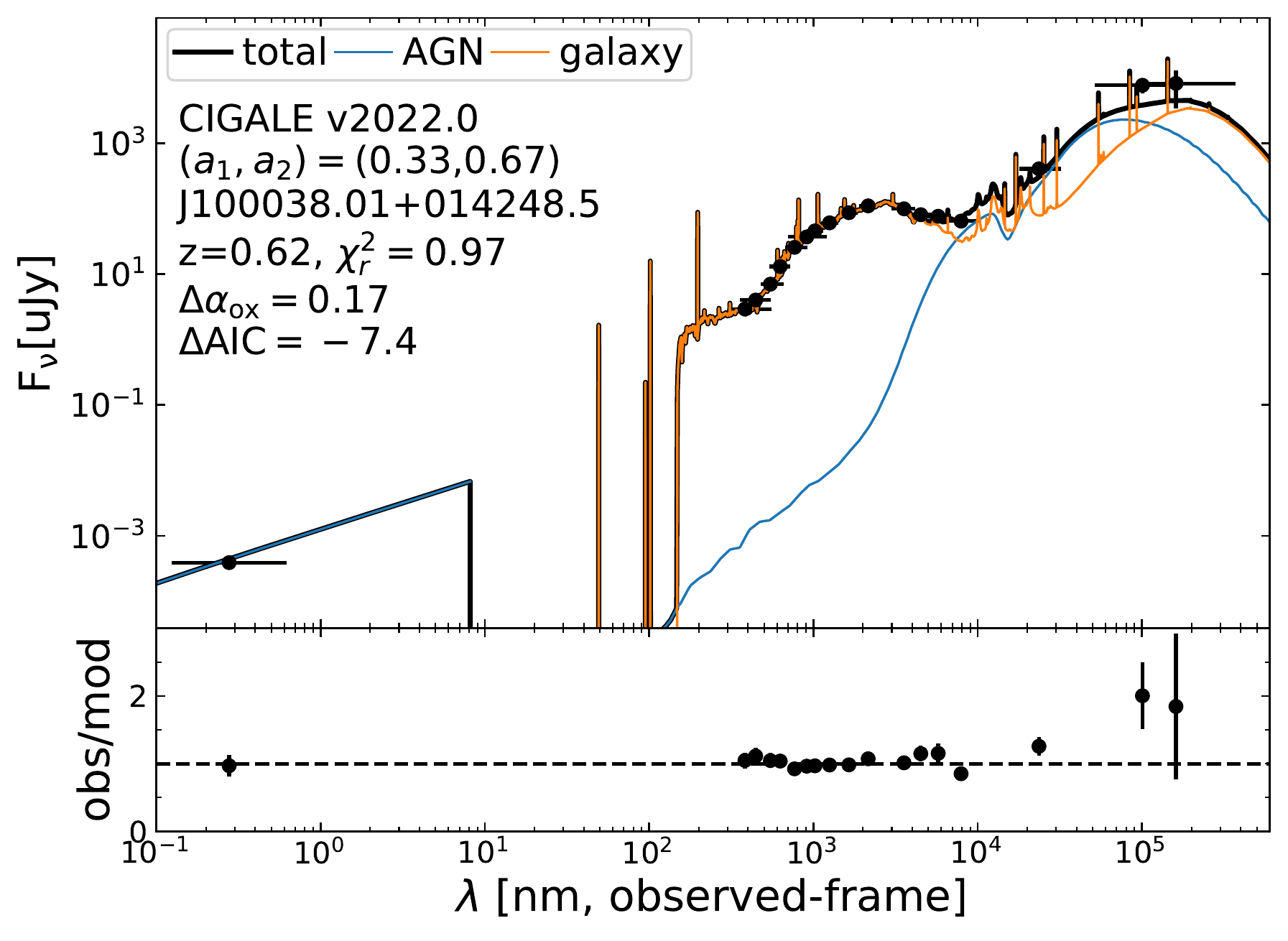}
    \caption{Example SED fits for a COSMOS type~2 AGNs from \xcig\ (top-left panel) and \cigv\ [three different $(a_1,a_2)$] as labeled.
    The \cigv\ fits have better quality than the \xcig\ fits,
    as indicated by the labeled $\Delta \rm AIC$ values (see \S\ref{sec:res_xray_ani} for the $\Delta \rm AIC$ definition).
    \fst{Therefore, the \xray\ anisotropic models are preferred over the isotropic one for this example source.}
    }
    \label{fig:cosmostype2_sed}
\end{figure*}

\begin{figure*}
    \centering
	\includegraphics[width=2\columnwidth]{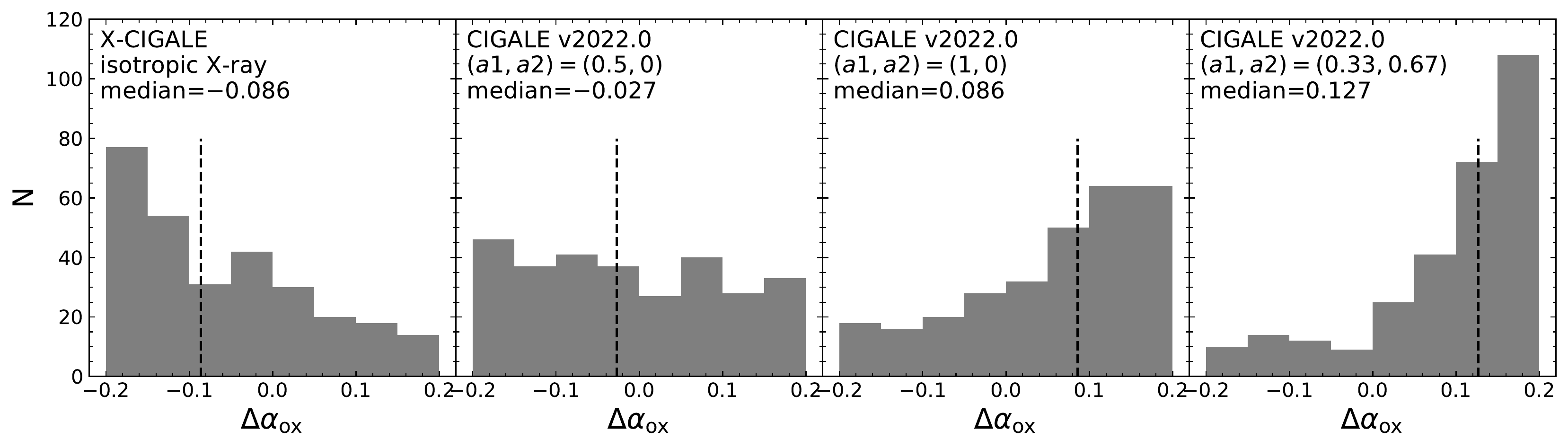}
    \caption{$\Delta \ox$ distributions for COSMOS type 2 AGNs from \xcig\ and \cigv\ [three different $(a_1,a_2)$] as labeled. 
    We use the best-fit $\Delta \ox$ values here, because \xcig\ cannot perform Bayesian-like analysis for the quantity of $\Delta \ox$ due to a technical reason. 
    The vertical dashed line indicates the median value of each distribution.
    These median values are also labeled on each panel. 
    The new \xcig\ fits with $(a_1, a_2)=(0.5, 0)$ have median $\Delta \ox$ closest to zero, indicating that this anisotropy model is the most physical among the four models (one isotropic and three anisotropic) tested. 
    }
    \label{fig:cosmostype2_det_ox}
\end{figure*}

In \xcig, the AGN \xray\ emission is modeled using the $\ox$-$\luvr$ relation from \cite{just07}, where $\luvr$ is the intrinsic disk emission at a viewing angle of 30$^\circ$\footnote{\fst{30$^\circ$ is the typical probability-weighted viewing angle for type~1 AGNs, assuming that the torus half-opening angle (between the equatorial plane and torus edge) is 40$^\circ$ (see \S2.2.3 of \citealt{yang20}).}} and $\ox$ is the AGN SED slope connecting $\luvr$ and $\lxr$.  
For type~1 AGNs, whose viewing angles are near 30$^\circ$ \fst{\citep{yang20}}, the SEDs are similar for isotropic and anisotropic \xray\ models in the framework of \xcig. 
However, for type~2 AGNs, whose viewing angles are much larger than 30$^\circ$, the isotropic and anisotropic models will predict significantly different \xray\ emissions, at a given AGN power. 

To test the effectiveness of \xcig\ (isotropic \xray\ emission), we use a spectroscopic type~2 AGN sample from the \chandra\ COSMOS-Legacy survey (\citealt{civano16, marchesi16}).
We require these AGNs to have $S/N>3$ in the hard band of \chandra\ (2--7~keV), and we apply absorption corrections to the hard-band fluxes based on the correction factors from \cite{marchesi16}, because \xcig\ requires that the input \xray\ fluxes are intrinsic \citep{yang20}.\footnote{\fst{For users without intrinsic \xray\ fluxes, it is feasible to estimate the absorption corrections on their own (see, e.g., \S3.1 of \citealt{mountrichas21}). To perform this task, users can first use {\sc behr} \citep{park06} to estimate the hardness ratios (HRs) based on hard and soft-band counts, which are often available in \xray\ catalogs. 
They can then input these HR values to {\sc pimms} \citep{mukai93} for the estimations of column density $N_H$ and intrinsic fluxes.
An alternative approach is to directly adopt the hard-band fluxes without absorption corrections, because hard \xray\ photons are penetrating and only modestly affected by absorption in general. 
For our case, the median correction for hard-band fluxes is only $\approx 5\%$. 
}}
The absorption corrections from \cite{marchesi16} are based on a standard hardness-ratio analyses.

We remove sources with $\lx < 10^{42.5}$~erg~s$^{-1}$, for which the \xray\ emission might originate from normal galaxies rather than AGNs \citep[e.g.,][]{aird17b}.  
We adopt the 14 broad-band photometric data ($u$ to IRAC~8~$\mu$m) from the COSMOS2015 catalog \citep{laigle16}.
We also use the MIPS~24~$\mu$m, PACS~100/160~$\mu$m, and SPIRE~250/350/500~$\mu$m photometry from the ``super-deblended'' catalog of \cite{jin18}.
There are a total of 296 type~2 AGNs, spanning a redshift range of 0.3--1.6 (10\%--90\% percentile).

Our fitting parameters are listed in Table~\ref{tab:par_type2}.
For the star formation history (SFH), we adopt a delayed $\tau$ SFH model and a \cite{bruzual03} simple stellar population model.
We adopt the \cite{calzetti00} galaxy attenuation law and the 
\cite{dale14} dust IR spectral templates.
For AGN IR emission, we adopt the SKIRTOR clumpy torus model 
\citep{stalevski12, stalevski16}.
\fst{We fix the torus half opening angle to the default 40$^\circ$, which is observationally preferred \citep[e.g.,][]{stalevski16}.
Under this setting, there are four type-2 viewing angles (60, 70, 80, and 90$^\circ$) available in SKIRTOR, and we allow all these values in our fits (see Table~\ref{tab:par_type2}).
The full SKIRTOR model set has another five parameters such as 9.7~$\mu$m optical depth and ratio of outer to inner radius.
These parameters generally have minor effects on the broad-band SED shapes (e.g., \citealt{yang20}), and thus we leave them at the default values to reduce the needed computing resources. 
In summary, we employ four templates (corresponding to different viewing angles) out of the total 19200 SKIRTOR models.}
In \xcig, AGN \xray\ and UV/optical emissions are related with the $\ox$-$\luvr$ relation of \cite{just07}, where $\ox$ is the UV/\xray\ slope calculated at the typical AGN type-1 viewing angle of $\theta=30^\circ$ \citep{yang20}, i.e.,
\begin{equation}
\label{eq:aox}
    \ox = -0.3838 \log \frac{\luvr (30^\circ)}{\lxr (30^\circ)},
\end{equation}
where $\luvr$ and $\lxr$ are the monochromatic AGN luminosities per frequency at rest-frame 2500~$\rm \AA$ and 2~keV, respectively.
\fst{Although the $\ox$-$\luvr$ relation is reasonably tight, it has a non-negligible intrinsic scatter of $\approx 0.1$ in terms of $\ox$ \citep[e.g.,][]{steffen06, just07}.
\xcig\ considers the scatter by constructing different models around the $\ox$-$\luvr$ relation, and the user can set the maximum deviation from the relation, $|\Delta \ox|_{\rm max}$
(see \S2.2.3 of \citealt{yang20} for details).
In our fits, we set $|\Delta \ox|_{\rm max}=0.2$ (Table~\ref{tab:par_type2}), about $2\sigma$ of the intrinsic scatter \citep[e.g.,][]{just07}.
We set the photon index $\Gamma=1.8$, a typical value for distant \xray\ AGNs \citep[e.g.,][]{yang16, liu17}.
}

Fig.~\ref{fig:cosmostype2_sed} (top-left) shows an example SED fit for one of the COSMOS type 2 AGNs.
Under the scheme of AGN unification \citep[e.g.,][]{antonucci93, urry95, netzer15}, our type~2 AGNs should also follow the $\ox$-$\luvr$ intrinsic relation (Eq.~\ref{eq:aox}).
To test this point, we plot the $\Delta \ox=\alpha_{\rm ox, fitted}-\alpha_{\rm ox, expected}$ (i.e., deviation from the $\ox$-$\luvr$ relation) distribution from our \xcig\ run in Fig.~\ref{fig:cosmostype2_det_ox}. 
The $\Delta \ox$ values tend to be systematically negative, with a median value of $-0.093$ (corresponding to a factor of 1.75 lower in terms of $\lxr/\luvr$) for \xcig. 
This result suggests that, with the assumption of isotropic AGN \xray\ emission in \xcig, the observed \xray\ fluxes of our type~2 AGNs tend to systematically lie below the expectations from the $\ox$-$\luvr$ relation. 
One natural solution to this issue is allowing intrinsic \xray\ anisotropy, so that an AGN viewed at type~2 angles will have lower \xray\ fluxes than viewed at type~1 angles.  
We perform this code-implementation task in \S\ref{sec:ani_code}.


\subsection{Code Improvement}
\label{sec:ani_code}
Considering the evidence for \xray\ anisotropy in \S\ref{sec:mot_xray_ani} and 
\S\ref{sec:ani_sample}, 
we modify the code so that the user can model $\lx$ as a 2nd-order polynomial function of the cosine of the viewing angle \citep[e.g.,][]{netzer87}:
\begin{equation}
\label{eq:a1a2}
    \frac{\lx (\theta)}{\lx (0)} = a_1 \cos\theta + a_2 \cos^2\theta + 1-a_1-a_2,
\end{equation}
where the coefficients ($a_1$ and $a_2$) are free parameters set by the user and $\theta$ is the viewing angle (face-on\ $= 0$, edge-on\ $=90^\circ$).
The constant term in Eq.~\ref{eq:a1a2}, $1-a_1-a_2$, guarantees that the right-hand side equals the left-hand side when $\theta=0$.
Setting $(a_1,a_2)=(0,0)$ mean isotropic $\lx$.
In \cigv, $\ox$ is still calculated using Eq.~\ref{eq:aox}.

\subsection{Results and interpretation}
\label{sec:res_xray_ani}
\begin{figure}
    \centering
    \includegraphics[width=\columnwidth]{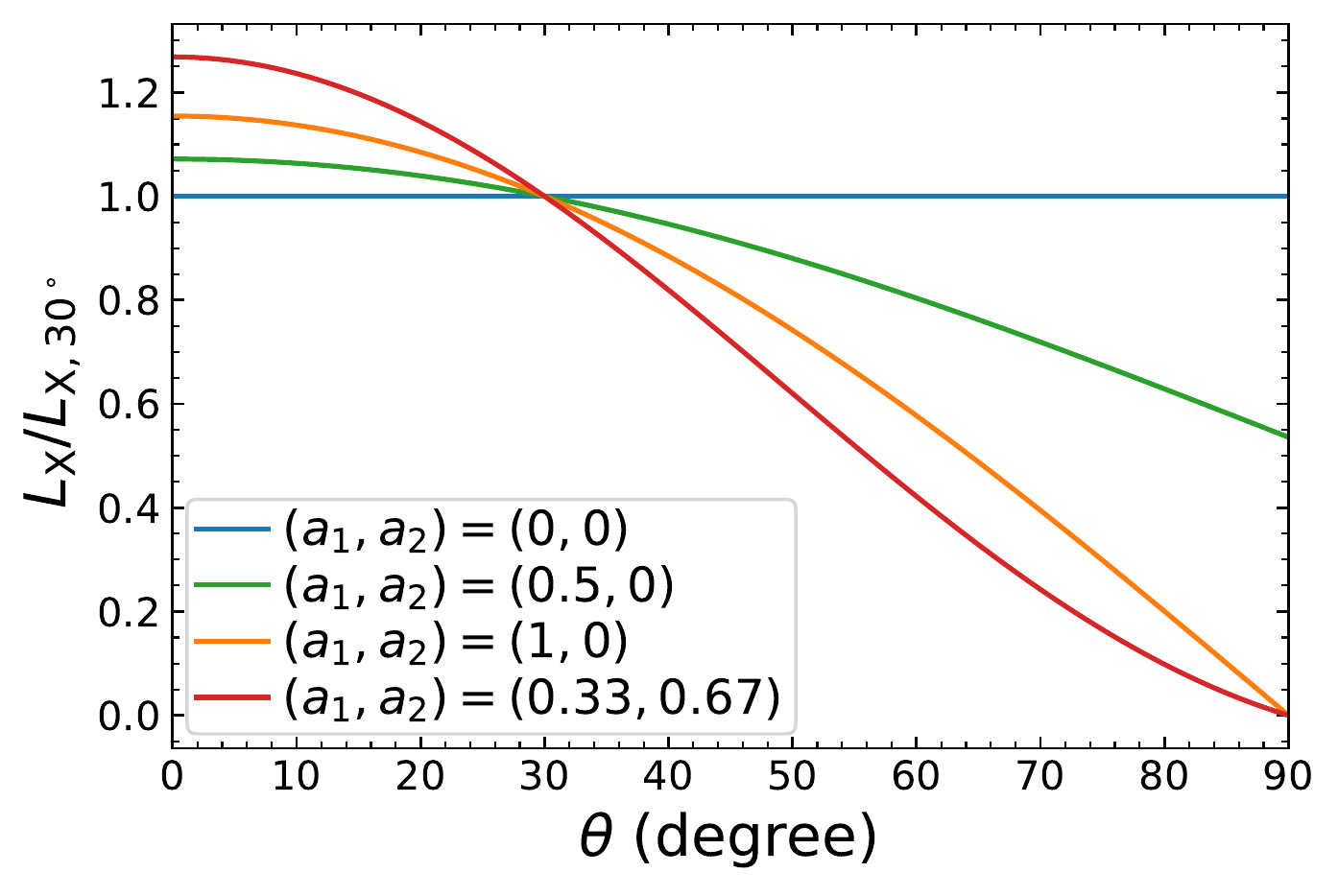}
    \caption{Dependence of $\lx$ on viewing angle.
    Different colors indicate different $(a_1,a_2)$ settings that are tested in this work.
    $(a_1, a_2)=(0,0)$ indicates the \xcig\ (isotropic) test.
    The $\lx$ ($y$-axis) is normalized at $\theta = 30^\circ$, where
    the $\ox$-$\luvr$ relation is applied in \xcig\ and \cigv. 
    }
    \label{fig:Lx_vs_theta}
\end{figure}

\begin{figure*}
    \centering
	\includegraphics[width=2\columnwidth]{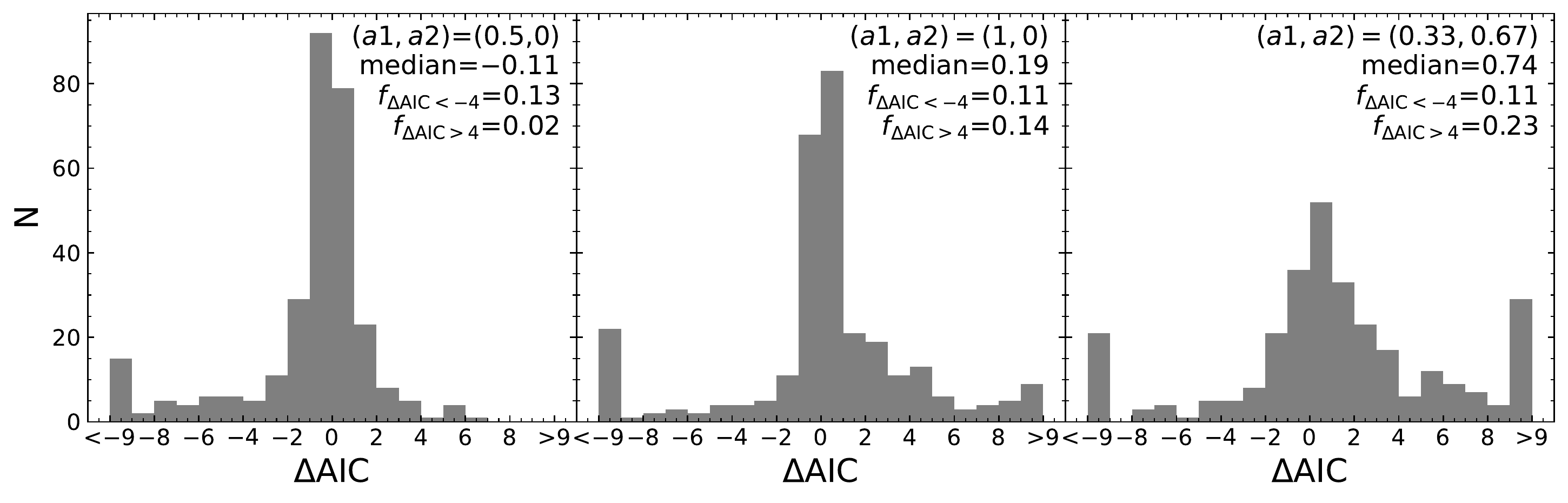}
    \caption{
    Distribution of AIC difference between the fits of \cigv\ and \xcig.
    Different panels are for the fits with different $(a_1, a_2)$ values as labeled. 
    The median value and the fractions of $\Delta \rm AIC<-4$ and $\Delta \rm AIC>4$
    are marked on each panel.
    The new \xcig\ fits with $(a_1, a_2)=(0.5, 0)$ has overall the best fit quality compared to other ones.  
    }
    \label{fig:cosmostype2_aic}
\end{figure*}

We repeat the fitting of type~2 AGNs (\S\ref{sec:mot_xray_ani}) but using \cigv. 
We perform three runs each with ($a_1$, $a_2$) set to $(0.5,0)$, $(1,0)$, and $(0.33,0.67)$, respectively, while other parameters are the same as the run in \S\ref{sec:ani_sample} (see Table~\ref{tab:par_type2}).
$(a_1,a_2)=(0.33,0.67)$ means the same angular dependence as that of the AGN disk emission ($\ld$) in the UV-to-IR AGN module (the SKIRTOR model; \citealt{stalevski12, stalevski16}).
$(a_1,a_2)=(1,0)$ is equivalent to a thin disk geometry with an angle-independent \xray\ intensity.  
The angle dependence of $(a_1,a_2)=(0.5,0)$ is between $(a_1,a_2)=(1,0)$ and the isotropic case.
Fig.~\ref{fig:Lx_vs_theta} displays the viewing angular dependence under these ($a_1$, $a_2$) settings. 
The angular dependence is stronger in the order of $(0.5,0)$, $(1,0)$, and $(0.33,0.67)$.

The SED fit for an example source is displayed in Fig.~\ref{fig:cosmostype2_sed}.
For this source, the anisotropic models have better fitting quality than the isotropic model (as indicated by the reduced $\chi^2$ labeled in Fig.~\ref{fig:cosmostype2_sed}).
Fig.~\ref{fig:cosmostype2_det_ox} displays the $\Delta \ox$ distributions from the \cigv\ runs. 
The settings of $(a_1, a_2)=(1, 0)$ and $(a_1, a_2)=(0.33, 0.67)$ lead to systematically positive $\Delta \ox$, 
with median values of 0.069 and 0.120, respectively. 
This result indicates that the angular dependence defined by these two parameter sets is overly strong. 
In contrast, the $\Delta \ox$ from $(a_1, a_2)=(0.5, 0)$ is more evenly distributed around zero compared to those from $(a_1, a_2)=(1, 0), (0.33, 0.67)$, and the \xcig\ (isotropic) result.
The $\Delta \ox$ median ($-0.027$) is the smallest among all four models, 
indicating that $(a_1, a_2)=(0.5, 0)$ is likely the most physical model among the tested ones.   

To assess the overall fitting quality, we calculate the difference of the Akaike information criterion ($\rm \Delta AIC$; \citealt{akaike74}) between the \cigv\ (anisotropic) and the \xcig\ (isotropic) fits.
This quantity is defined as $\Delta \rm AIC=2\Delta k+\Delta \chi^2$,
where $\Delta k$ is the difference in the number of free parameters.
$\Delta k$ is zero for our case here. 
Lower $\Delta \rm AIC$ indicates stronger probability of anisotropic models.
For example, $\Delta \rm AIC<-4$ means the anisotropic model is more than $\approx 7$ ($e^{-\Delta \rm AIC/2}$) times more probable than the isotropic model, indicating a strong support for the former \citep[e.g.,][]{burnham02}.

The example source in Fig.~\ref{fig:cosmostype2_sed} has $\Delta \rm AIC<-4$, indicating that the anisotropic models are preferred over the isotropic model. 
When inspecting the residuals in Fig.~\ref{fig:cosmostype2_sed}, one could be puzzled that the main difference between \xcig\ vs.\ \cigv\ fits is in the IR rather than \xray. 
We note that the root of this difference is not related to the IR AGN emission model, as all fits are based on the same IR AGN models (Table 1). 
Instead, the actual cause is \xray\ angle dependence, which is the only different setting among the fits.
\scd{We briefly explain this cause below.}

\scd{For our COSMOS type~2 sample, the AGNs have emission mostly in \xray\ and IR as their UV/optical radiation is obscured. 
The \xray/IR ratio is an observable quantity closely related to the \xray\ angle dependence \citep[e.g.,][]{asmus15}. 
To see this point, we can write the \xray/IR ratio as 
\begin{equation}
\label{eq:xir}
    \frac{\lx(\theta)}{\lir(\theta)} = \frac{\lx(\theta)}{\lx(30^\circ)} \times
                                       \frac{\lx(30^\circ)}{\luvo(30^\circ)} \times
                                       \frac{\luvo(30^\circ)}{\lir(\theta)},
\end{equation}
where $\lx$, $\lir$, and $\luvo$ are AGN \xray, IR, and intrinsic UV/optical luminosities, respectively.
In Eq.~\ref{eq:xir}, the second factor $\frac{\lx(30^\circ)}{\luvo(30^\circ)}$ is roughly a constant value, as \cig\ directly links AGN \xray\ and UV/optical emission by the $\ox$-$\luvr$ relation at a $30^\circ$ viewing angle \citep{yang20}.
The third factor $\frac{\luvo(30^\circ)}{\lir(\theta)}$ is also about a constant (depending on the dust-model details), because the IR emission originates from the UV/optical photons absorbed by dust and \cig\ strictly keeps energy conservation \citep{boquien19, yang20}.
Therefore, the \xray/IR ratio is approximately proportional to the first factor $\frac{\lx(\theta)}{\lx(30^\circ)}$, which is the \xray\ angle dependence. 
For type~2 viewing angles, compared to the \xray\ isotropic model, our tested anisotropic models have lower $\frac{\lx(\theta)}{\lx(30^\circ)}$ values  (Fig.~\ref{fig:Lx_vs_theta}) and thereby systematically lower \xray/IR ratios. 
For the source in Fig.~\ref{fig:cosmostype2_sed}, the observed IR/\xray\ ratio is more similar to the anisotropic model values than the isotropic one.
This is the reason why the anisotropic configurations model the observed data better than the isotropic one.}

\scd{This source in Fig.~\ref{fig:cosmostype2_sed} is also a representative example demonstrating that different bands are not modeled independently. 
\cig\ templates are rigid across all bands, finding a solution that minimizes the ``global'' $\chi^2$, although such a solution might not minimize residuals in some bands.}

Fig.~\ref{fig:cosmostype2_aic} shows the distribution of $\Delta \rm AIC$.
For the model of $(a_1, a_2)=(0.5, 0)$, 13\% of the sources have $\Delta \rm AIC<-4$, while only 2\% sources have $\Delta \rm AIC>4$.
The overall distribution is towards the negative sign (median $\Delta \rm AIC=-0.11$).
This result indicates that the fitting quality has an overall improvement from the isotropic model to the $(a_1, a_2)=(0.5, 0)$ anisotropic model.
However, from Fig.~\ref{fig:cosmostype2_aic}, the models of $(a_1, a_2)=(1, 0)$ and $(0.33, 0.67)$ have similar or even worse fitting quality than that of the isotropic model.  

From the $\Delta \ox$ and the AIC analyses above, an isotropic AGN \xray\ model is disfavored compared to the anisotropic model with $(a_1, a_2)=(0.5, 0)$. 
Therefore, AGN \xray\ emission is likely weaker toward larger viewing angles, qualitatively consistent with the observations of \citet[][see \S\ref{sec:mot_xray_ani}]{liu14, asmus15}. 
On the other hand, the amplitude of this viewing-angle dependence is moderate, since $(a_1, a_2)=(0.5, 0)$ results in better fitting quality than $(a_1, a_2)=(1, 0)$ and $(0.33, 0.67)$, which have stronger angular dependence (see Fig.~\ref{fig:Lx_vs_theta}). 
The conclusion that AGN X-rays have weaker angular dependence than UV/optical [$(a_1, a_2)=(0.33, 0.67)$] is understandable. 
The \xray\ photons result from the inverse Compton scattering of the UV/optical seed photons, and the strength of anisotropy is suppressed by this scattering process \citep[e.g.,][]{xu15b, yang20}.

We caution that our conclusion of \xray\ angle dependence is for the overall \fst{type~2} AGN population rather than individual sources, as our analyses above are based on the statistical analyses of the entire \fst{type~2} sample.
It might be possible that individual AGNs have different angular dependence, because the structure of the AGN corona, which produces the \xray\ photons (\S\ref{sec:mot_xray_ani}), could vary among individual sources \citep[e.g.,][]{ricci18, tortosa18}.  

\fst{We set $\Gamma=1.8$ in our runs (Table.~\ref{tab:par_type2}), but the actual power-law photon index for an \xray\ AGN may range from $\approx 1.6$ to $\approx 2.2$ \citep[e.g.,][]{yang16, liu17}.
The photon-index parameter can affect the model-predicted \xray\ fluxes. 
To assess this effect, we repeat our runs allowing $\Gamma$ varying between 1.6 and 2.0.
The resulting $\Delta \ox$ and $\Delta \rm AIC$ distributions are similar to those in Figs.~\ref{fig:cosmostype2_det_ox} and \ref{fig:cosmostype2_aic}, and the $(a_1,a_2)=(0.5,0)$ configuration is still the most favored model. 
In Table.~\ref{tab:par_type2}, we adopt large viewing angles ($\geq 60^\circ$) assuming the classic unification model, i.e., type~2 AGNs are obscured by the torus. 
However, some recent observations suggest that type~2 AGNs might also have small viewing angles and be obscured by polar dust \citep[e.g.,][]{mountrichas21c, ramos21}.
To consider this possibility, we test new \cig\ runs allowing all available viewing-angle values in SKIRTOR (0--90$^\circ$ with a step of 10$^\circ$).
The result still favors the $(a_1,a_2)=(0.5,0)$ anisotropic model, consistent with our original result.
Based on the tests above, we consider our main conclusion not to be critically dependent on the adopted parameters of the photon index and viewing angle in Table~\ref{tab:par_type2}.
}

\fst{In the code of \cigv}, we set the default $(a_1, a_2)$ to $(0.5,0)$ based on our results above.
For general purposes of AGN modelling, the user does not need to change these default values. 
For the specific purposes of studying AGN \xray\ anisotropy, the user can test different $(a_1, a_2)$ values in different runs and select the best parameters, like our approach above. 
This method allows further studies of \xray\ angular dependence for different AGN samples (e.g., high-accretion rates versus low-accretion rates), and thereby can provide insight into the properties of AGN corona.

\section{Normal-galaxy X-ray Emission}
\label{sec:xrb}
\subsection{Motivation}
Both AGNs and normal galaxies can emit \xray\ photons. 
Normal-galaxy X-rays originate primarily from point sources of \xray\ binaries and diffuse hot gas. 
AGNs tend to be more luminous than normal galaxies at \xray\ wavelengths. 
As a consequence, most of the \xray\ detected sources in extragalactic surveys are AGNs.
However, normal galaxies become increasingly important as survey depth improves. 
The 7~Ms \chandra\ Deep Field-South (CDF-S) survey has $\approx 30\%$ of the \xray\ detections classified as normal galaxies and such sources dominate the faintest detections \citep{luo17}.
It is thereby expected that many more normal galaxies will be detected in deep surveys by future \xray\ telescopes with large collecting areas such as \athena\ and \lynx. 
Therefore, it is critical to have realistic recipes for normal-galaxy \xray\ modeling. 

{\sc x-cigale} has both AGN and galaxy \xray\ components \citep{yang20}.
The latter includes the emission from high-mass \xray\ binaries (HMXBs), low-mass \xray\ binaries (LMXBs), and hot gas. 
The AGN component has been well tested \citep[e.g.,][]{yang20, zou20, mountrichas21}, but this is not the case for the galaxy component.
Below, we test and improve the modeling of galaxy \xray\ emission. 

\subsection{Sample and preliminary fitting}
\label{sec:mot_xrb}

\begin{table*}
\centering
\caption{Model parameters for the CDF-S normal galaxies}
\label{tab:par_normal}
\begin{tabular}{llll} \hline\hline
Module & Parameter & Symbol & Values \\
\hline
\multirow{2}{*}{\shortstack[l]{Star formation history\\
                               $\mathrm{SFR}\propto t \exp(-t/\tau)$ }}
    & Stellar e-folding time & $\tau_{\rm star}$ & 0.1, 0.5, 1, 5 Gyr\\
    & Stellar age & $t_{\rm star}$  
            & 0.5, 1, 3, 5, 7 Gyr\\ 
\hline
\multirow{2}{*}{\shortstack[l]{Simple stellar population\\ 
    \cite{bruzual03}}}
    & Initial mass function & $-$ & \cite{chabrier03} \\
    & Metallicity & $Z$ & 0.004, 0.02 \\
\hline
\multirow{2}{*}{\shortstack[l]{Dust attenuation \\ 
                \cite{calzetti00} }}
    & \multirow{2}{*}{Color excess of the nebular lines} & 
        \multirow{2}{*}{$E(B-V)$} &
        \multirow{2}{*}{\shortstack[l]{0.05, 0.1, 0.2, \\
                                    0.3, 0.4, 0.5, 0.6 mag}} \\\\
\hline
\multirow{2}{*}{\shortstack[l]{Galactic dust emission: \\ \cite{dale14}}}
    & \multirow{2}{*}{\shortstack[l]{Slope in $dM_{\rm dust} \propto U^{-\alpha} dU$}}
    & \multirow{2}{*}{\shortstack[l]{$\alpha$}}
    & \multirow{2}{*}{\shortstack[l]{2}}
    \\\\
\hline
\multirow{3}{*}{\shortstack[l]{AGN (UV-to-IR) \\ SKIRTOR }}
    & AGN contribution to IR luminosity & $\fracA$ & 0, 0.01, 0.03, 0.1, 0.2  \\
    & Viewing angle & $\theta$ & 30$^\circ$, 70$^\circ$ \\
    & \multirow{1}{*}{\shortstack[l]{Polar-dust color excess}} & \multirow{1}{*}{\shortstack[l]{$E(B-V)_{\rm PD}$}} & \multirow{1}{*}{\shortstack[l]{0}} \\
\hline
    \multirow{2}{*}{\shortstack[l]{X-ray}}
    & \textbf{Deviation from the expected} $\boldsymbol{\log \lhmxb}$  & $\boldsymbol{\dethmxb}$ & 
        \textbf{$\boldsymbol{-0.5}$ to 0.5 (step 0.1) dex} \\
    & \textbf{Deviation from the expected} $\boldsymbol{\log \llmxb}$  & $\boldsymbol{\detlmxb}$ & 
        \textbf{$\boldsymbol{-0.5}$ to 0.5 (step 0.1) dex} \\
\hline
\end{tabular}
\begin{flushleft}
{\sc Note.} --- For parameters not listed here, we use the default values.
\textbf{Bold font} indicates new parameters in \cigv\ introduced in this work.
\end{flushleft}
\end{table*}

\begin{figure*}
    \centering
	\includegraphics[width=2\columnwidth]{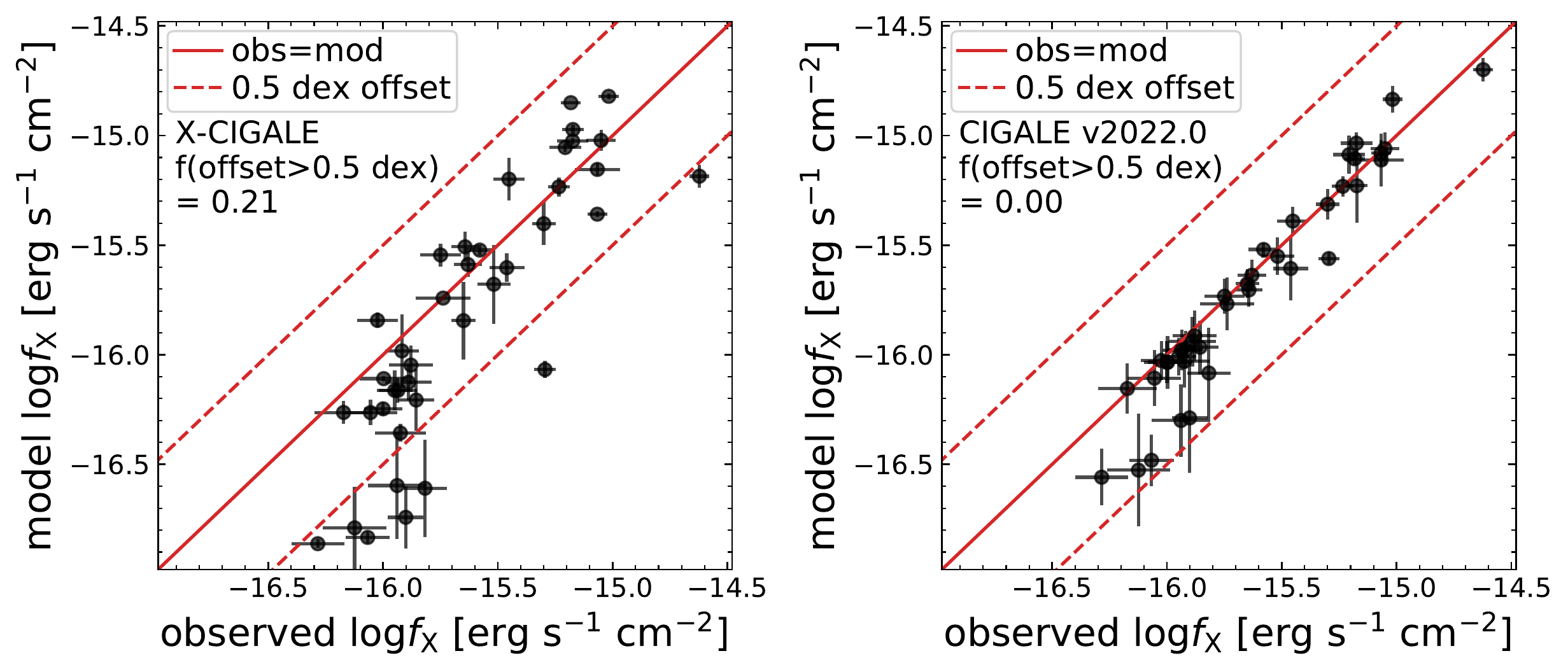}
    \caption{Model versus\ observed \xray\ 0.5--7~keV flux for our CDF-S normal galaxies from the \xcig\ (left) and \cigv\ (right).
    The red solid line indicates a model$=$observed relation; the red dashed lines indicate 0.5~dex offsets from this relation.
    }
    \label{fig:normal_obs_vs_mod}
\end{figure*}

\begin{figure*}
    \centering
	\includegraphics[width=\columnwidth]{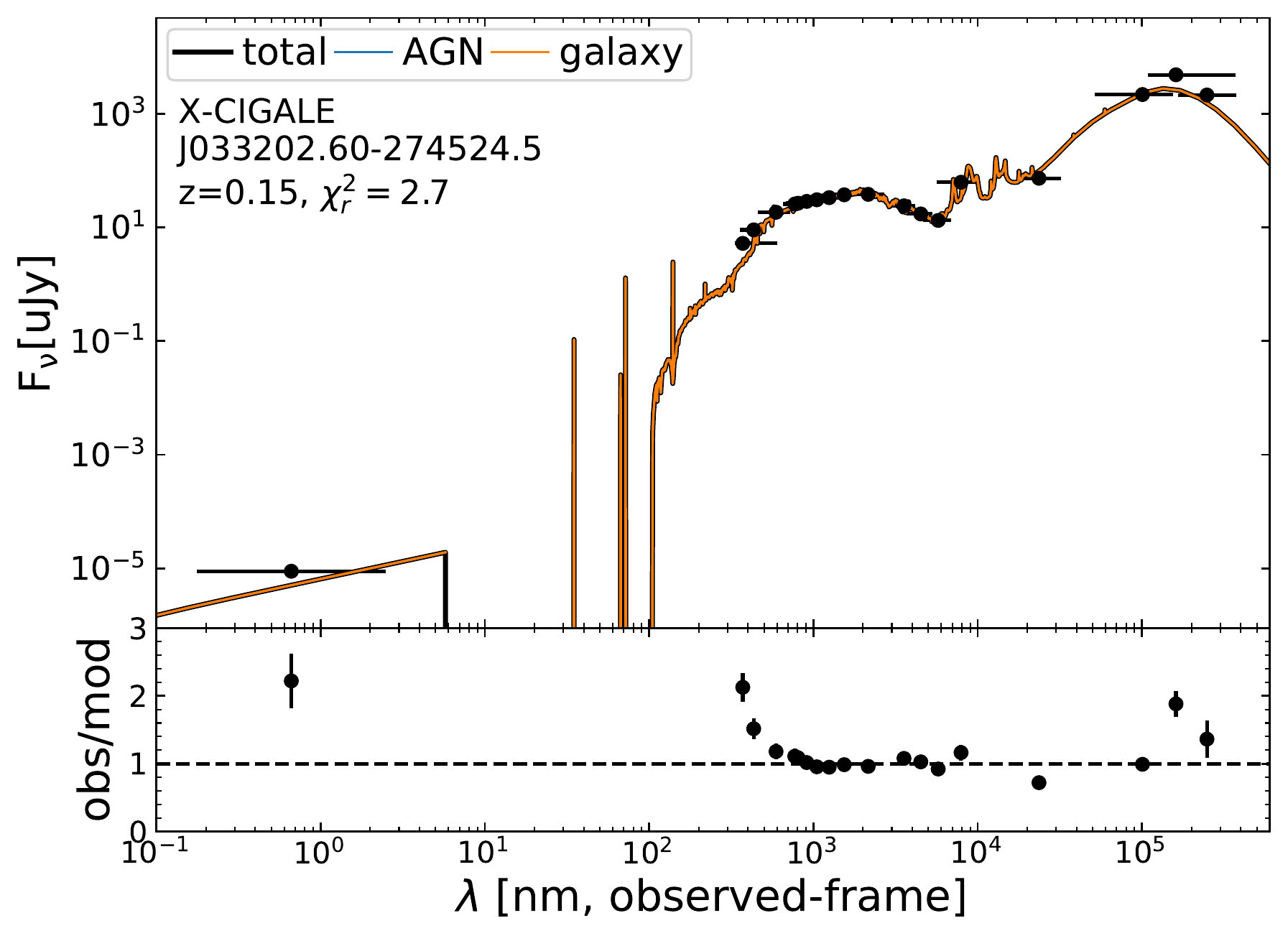}
	\includegraphics[width=\columnwidth]{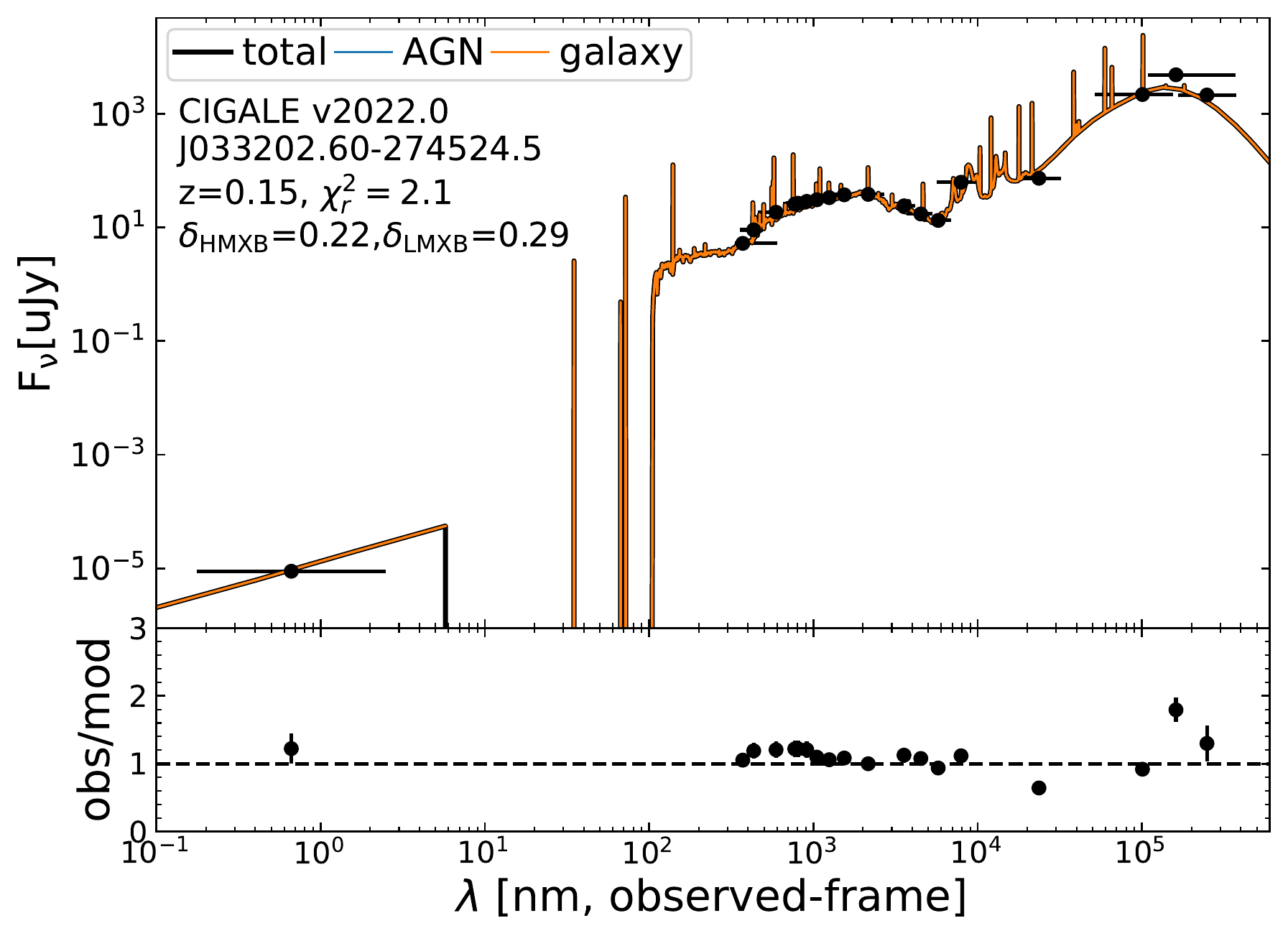}
    \caption{Example SED fits for a CDF-S normal galaxy from \xcig\ (left) and \cigv\ (right).
    \fst{The best-fit $\fracA$ is zero in both fits.}
    The observed 0.5--7~keV \xray\ flux is much higher than the model one from the \xcig, but is similar to the one from \cigv.
    Also, we note that \cigv\ also has a better fit to the UV data than \xcig.
    This is because \xcig\ is forced to use a stellar population model that corresponds to a relatively high \xray\ flux, although this model does not well fit the observed UV fluxes. 
    }
    \label{fig:normal_sed}
\end{figure*}

We test the galaxy \xray\ modeling in \xcig\ using the 7~Ms CDF-S survey, which is the deepest \xray\ survey to date \citep{luo17}. 
We take advantage of this unique dataset to study the \xray\ emission 
from normal galaxies in the distant universe, as galaxies' \xray\ 
power is typically low ($\lx\lesssim 10^{42.5}$~erg~s$^{-1}$) and 
beyond the sensitivity of most \xray\ surveys. 

We first select sources classified as ``galaxy'' instead of ``AGN'' or 
``star'' by \cite{luo17}. 
The classification is based on \xray\ and other multiwavelength data.  
We then restrict the sample to only sources within the GOODS-S field \citep{guo13}, where deep multiwavelength coverage is available from
UV to FIR. 
We compile the UV-to-IRAC4 data from \cite{guo13} and \spitzer/\herschel\ 
mid-to-far IR data from the ASTRODEEP team (Tao Wang 2020, private 
communication, GOODS-S \herschel\ catalog).
We discard sources with MIPS~24~$\mu$m S/N~$<3$, as reliable IR data are essential in constraining SFR (which scales with $\lhmxb$) and possible low-level AGN activity.  
There are a total of 39 \xray\ detected galaxies in our sample. 
We adopt the redshift measurements compiled by \cite{luo17}, which are 
secure spectroscopic redshifts or high-quality photometric redshifts.  
The redshifts cover a range of $z=0.10$--1.06 (10\%--90\% percentile), 
with a median of $z=0.44$. 

We first perform SED modeling of these galaxies using \xcig.
The fitting parameters are summarized in Table~\ref{tab:par_normal}.
The galaxy settings are similar to those in \S\ref{sec:xray_ani}, 
except that we allow two metallicity values of $Z=0.004$ and 0.02, as the
$\lhmxb$-SFR scaling relation depends on metallicity \citep[e.g.,][]{fragos13b}.
We still allow a moderate AGN component in the fitting ($\fracA\leq 0.2$). 
Although the sources are classified as galaxies by \cite{luo17}, some could possibly be low-luminosity AGNs \citep[e.g.,][]{young12, ding18}.

We compare the model versus\ observed \xray\ fluxes in Fig.~\ref{fig:normal_obs_vs_mod} for \xcig\ (left panel).
For many (21\%) of our sources, the offsets between the model and observed fluxes are more than 0.5~dex. 
We show an example SED fit with such an issue in Fig.~\ref{fig:normal_sed} (left). 
Therefore, \xcig\ is not able to model well all of the observed \xray\ fluxes.

\subsection{Code Improvement}
\label{sec:normal_code}
{\sc x-cigale} assumes that galaxy \xray\ emission from HMXBs and LMXBs can be calculated from the scaling relations of $\lhmxb$-SFR and $\llmxb$-$\mstar$ \citep[][]{fragos13b}.
However, this is an oversimplified assumption, because these relations are just an approximation for the overall galaxy population and scatters around them exist. 
For example, the content of globular clusters at a given $\mstar$, which is not modeled in \xcig, can significantly affect $\llmxb$ \citep[e.g.,][]{lehmer20}. 
Also, since the HMXB and LMXB emissions are from discrete point sources, $\lhmxb$ and $\llmxb$ inevitably suffer from statistical fluctuations that are especially strong in low-SFR and/or low-$\mstar$ galaxies \citep[e.g.,][]{lehmer19,lehmer21}.

To model the $\lhmxb$ and $\llmxb$ dispersions of individual galaxies in better detail, we introduce two new free parameters, $\dethmxb$ and $\detlmxb$, to account for the scatters of the $\lhmxb$-SFR and $\llmxb$-$\mstar$ scaling relations, i.e., 
\begin{equation}
\begin{split}
\log(\frac{L_{\rm 2-10\ keV}^{\rm HMXB}}{\rm SFR}) = 40.3 - 62Z + 569Z^2 - 1834Z^3 \\ 
            + 1968Z^4 + \dethmxb \\
\log(\frac{L_{\rm 2-10\ keV}^{\rm LMXB}}{\mstar}) = 40.3 - 1.5\log t - 0.42(\log t)^2 \\ 
            + 0.43(\log t)^3 + 0.14(\log t)^4 + \detlmxb \\
\end{split}
\end{equation}
where $\mstar$ and SFR are in solar units; $t$ denotes stellar age
in units of Gyr; $Z$ denotes metallicity (mass fraction).
The parameters of $\dethmxb$ and $\detlmxb$ are logarithmic 
deviations from the scaling relations, with positive/negative 
values meaning higher/lower HMXB and LMXB luminosities, respectively. 
The user can set multiple values for each parameter to enable a more flexible XRB prescription. 

Besides introducing $\dethmxb$ and $\detlmxb$, we also implement another update of the code. 
The code provides three SFR parameters: the instantaneous SFR, the average SFR over 10 Myr, and the average SFR over 100 Myr.
While \xcig\ adopted the instantaneous SFR when calculating $\lhmxb$, 
we adopt the average SFR over 100 Myr in \cigv. 
This change is because the HMXB emission has strong variability 
on $\sim 10$~Myr timescales \citep[e.g.,][]{linden10,garofali18,antoniou19}, 
but we do not have well-informed calibrations for how the $\lhmxb$ 
varies on such short timescales. 
On the other hand, the $\lhmxb$ dependence on longer timescales 
of $\sim 100$~Myr has been carefully characterized  
\citep[e.g.,][]{lehmer19, lehmer21}.
Although the instantaneous SFR and the 100-Myr averaged SFR are similar for a smooth star-formation history (SFH), they can differ significantly if a recent burst/quenching is present in the SFH.

\subsection{Results and interpretation}
\label{sec:res_xrb}

\begin{figure}
    \centering
	\includegraphics[width=\columnwidth]{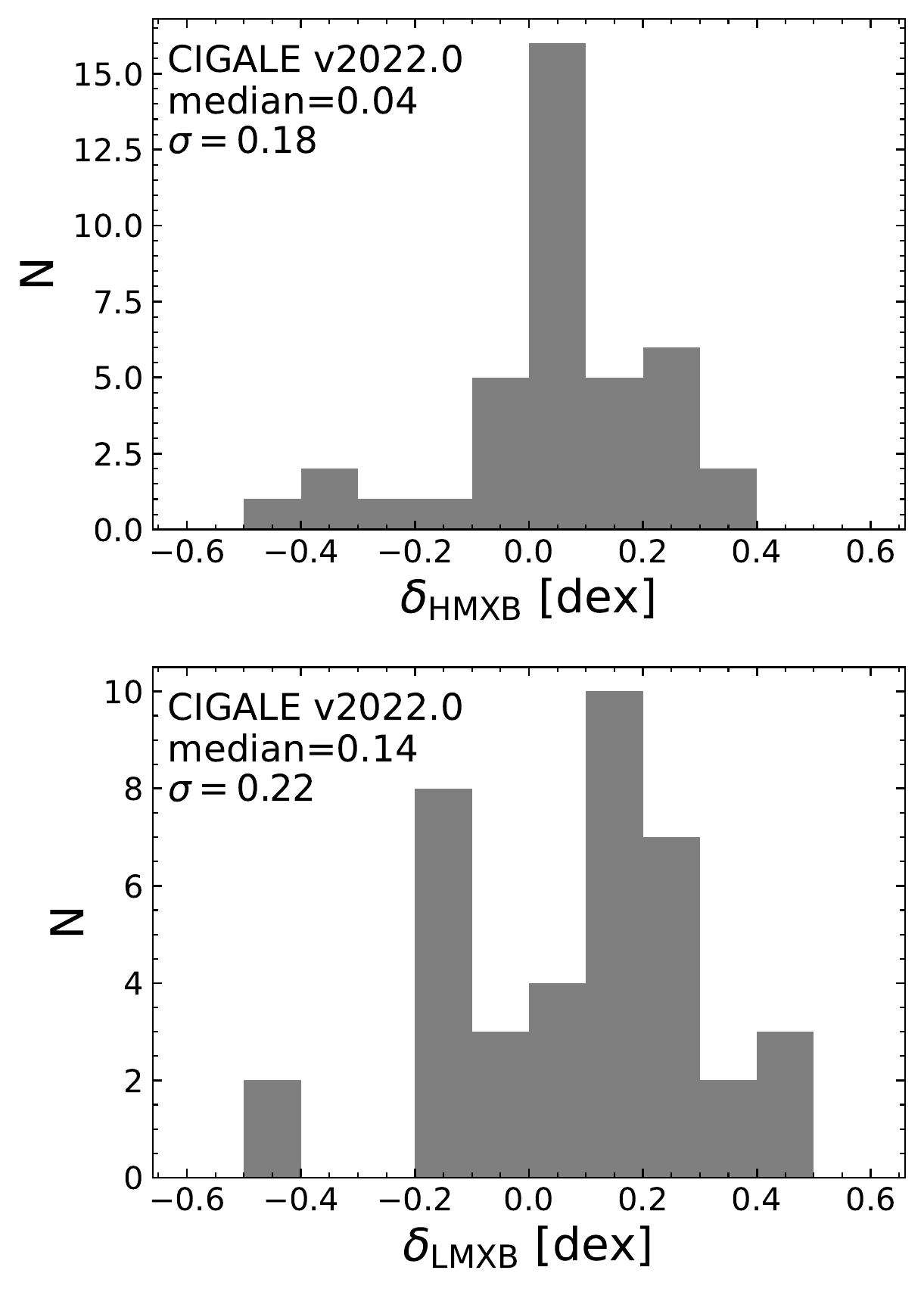}
    \caption{The distributions of $\dethmxb$ (top) and $\detlmxb$ (bottom) for CDF-S normal galaxies from the fits of \cigv.
    The median and standard deviation are labeled on each panel. 
    The medians are close to zero, indicating that the $\lhmxb$ and $\llmxb$ scaling relations \citep[][]{fragos13b} are good approximations for the overall galaxy population detected in deep \xray\ surveys at $z\lesssim 1$.
    }
    \label{fig:normal_det_hist}
\end{figure}

\begin{figure}
    \centering
	\includegraphics[width=\columnwidth]{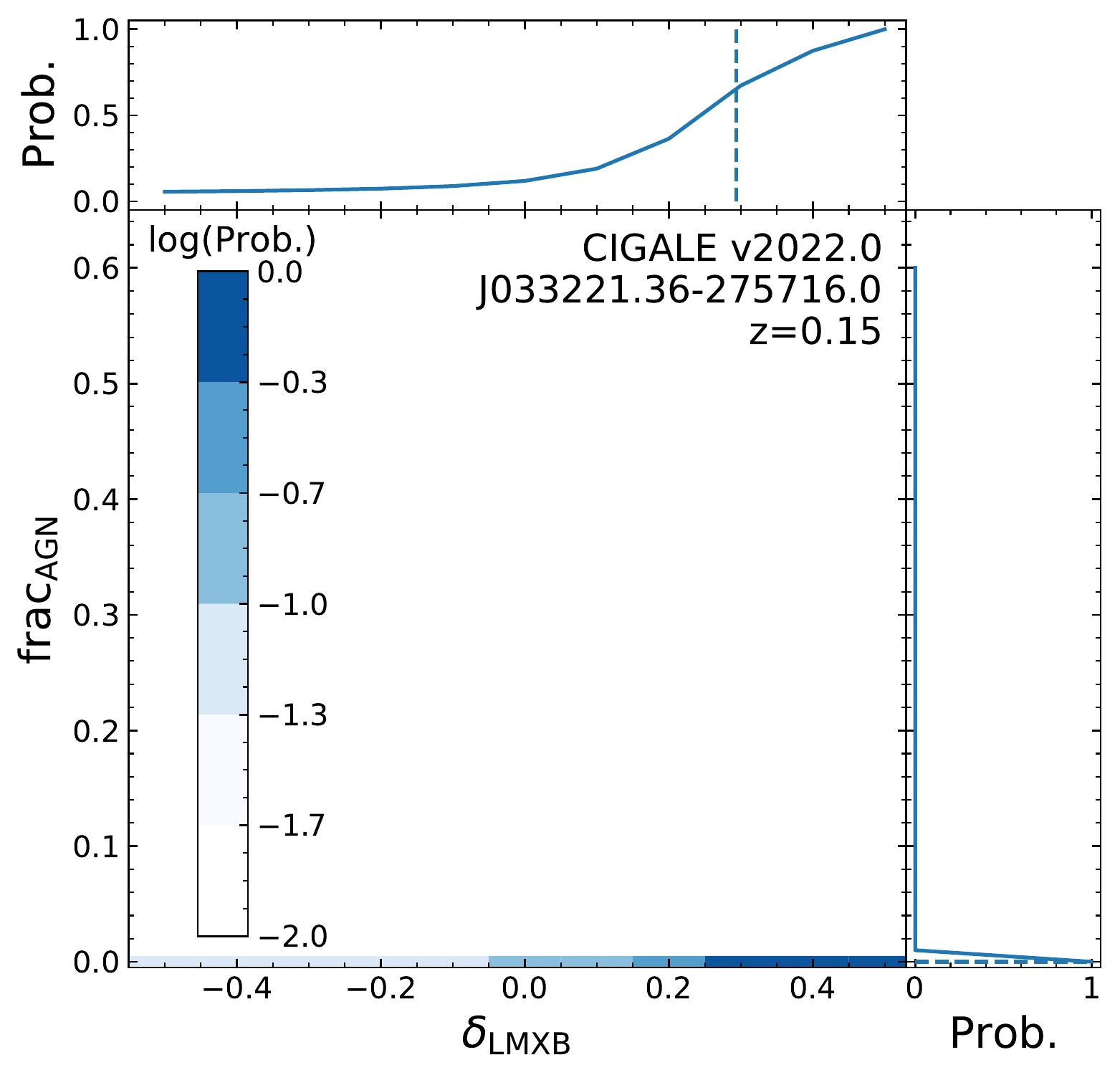}
    \caption{\fst{2D and 1D probability distributions of $\fracA$ and $\detlmxb$ for the CDF-S normal galaxy in Fig.~\ref{fig:normal_sed}.
    The plot is generated from the \cigv\ run with $\fracA=0$--0.6.
    The dashed line indicates the Bayesian (probability-weighted average) values of the two parameters. 
    The $\fracA$ is tightly constrained at a low level ($\lesssim 0.01$) despite the wide allowed $\fracA$ range in the fit.
    The situations are similar for other sources.}
    }
    \label{fig:normal_2d_prob}
\end{figure}

With \cigv, we re-fit our sample of CDF-S normal galaxies.
We set $\dethmxb$ from $-0.5$ to 0.5~dex with a step
of 0.1~dex, while keeping the other parameters unchanged 
(Table~\ref{tab:par_normal}). 
This parameter range is chosen because \cite{lehmer21} found the 2$\sigma$ scatter of $\lhmxb/$SFR is $\approx 0.5$~dex at SFR $\approx 4 M_\odot$~yr$^{-1}$, which is the median SFR of our sample. 
We also set $\detlmxb$ to the same values as $\dethmxb$. 

Fig.~\ref{fig:normal_obs_vs_mod} (right) compares the \cigv\ resulting model versus observed \xray\ fluxes. 
The model fluxes agree much better with the observed fluxes compared to those fitted by \xcig\ (Fig.~\ref{fig:normal_obs_vs_mod} left).
The offsets between the model and observed are all within 0.5~dex. 
Fig.~\ref{fig:normal_sed} (right) shows an example SED fit with \cigv. 
For this example, compared to \xcig, \cigv\ has a better fit not only to the \xray\ data but also to the UV data. 
This is because \xcig\ is forced to use a stellar population model that corresponds to a relatively high \xray\ emission, although this model does not well fit the observed UV fluxes.
This example highlights the importance of introducing $\dethmxb$ and $\detlmxb$, without which inappropriate stellar models might be selected. 

Fig.~\ref{fig:normal_det_hist} displays the distributions of the fitted $\dethmxb$ and $\detlmxb$, respectively. 
Both distributions have slightly positive median values, i.e., 0.04~dex (HMXB) and 0.12~dex (LMXB). 
These near-zero medians indicate that the $\lhmxb$ and $\llmxb$ scaling relations \citep[][]{fragos13b} are good approximations for the overall galaxy population detected in deep \xray\ surveys at $z\lesssim 1$. 
The slightly positive trend of the distributions suggest that the scaling relations might have systematic offsets. 
But the positive trend is expected due to a selection effect, because our \xray\ data are flux-limited and thus tend to select higher $\lhmxb$ and $\llmxb$ sources.
A larger normal-galaxy \xray\ sample, from, e.g., \textit{eROSITA} \citep[e.g.,][]{vulic21}, is needed to investigate the nature the positive trend of $\dethmxb$ and $\detlmxb$.
Both of the $\dethmxb$ and $\detlmxb$ distributions have substantial scatters (standard deviations $\approx 0.2$~dex; Fig.~\ref{fig:normal_det_hist}).
These scatters are likely caused by, e.g., globular-cluster contents and statistical fluctuations (see \ref{sec:normal_code}).

\fst{We set $\fracA\leq 0.2$ in our runs (Table~\ref{tab:par_normal}) since the sources were classified as normal galaxies by \cite{luo17}, but this quantitative choice is rather arbitrary. 
One might worry that our fitting results could heavily depend on the assumption of $\fracA\leq 0.2$, as AGNs can also contribute to the observed \xray\ fluxes. 
To assess the effects of this assumption, we perform a new run including higher $\fracA$ values, i.e., $\fracA=0, 0.01, 0.03, 0.1, 0.2, 0.3, 0.4, 0.5, 0.6$,
while keeping other parameters the same.
The resulting $\dethmxb$ and $\detlmxb$ distributions are identical to those in Fig.~\ref{fig:normal_det_hist}, indicating that these two parameters are not strongly degenerate with $\fracA$. 
Fig.~\ref{fig:normal_2d_prob} displays example 2D and 1D probability distributions of $\fracA$ and $\detlmxb$ from the new fit with $\fracA=$0--0.6.
The $\fracA$ is tightly constrained at a low level ($\lesssim 0.01$), and the situations are similar for all of our sources. 
The tight AGN constraints are understandable as the AGN emission is constrained not only by the \xray\ data but also by the UV-to-IR data. 
In summary, we conclude that our fitting results based on the parameters in Table~2 do not depend on the assumption of $\fracA\leq 0.2$. 
}

We set the default values of $\dethmxb$ and $\detlmxb$ both to 0, corresponding to the standard \cite{fragos13} scaling relations. 
For luminous \xray\ sources (e.g., $\lx \gtrsim 10^{42.5}$~erg~s$^{-1}$), the observed \xray\ fluxes are likely dominated by AGNs, and thus the user can just keep the default values of $\dethmxb=\detlmxb=0$.
For less luminous sources, galaxy \xray\ emission could dominate the observed fluxes, and the user can adopt different values of $\dethmxb$ and $\detlmxb$ (e.g., Table~\ref{tab:par_normal}) to allow more flexible XRB modeling. 
For specific galaxy populations, the user could allow multiple values for only one of $\dethmxb$ and $\detlmxb$ to save memory and reduce computation time. 
For example, for quiescent galaxies, $\lhmxb$ should be negligible compared to $\llmxb$. 
In this case, the user can adopt multiple values for $\detlmxb$ while keeping $\dethmxb=0$. 

\section{Flexible UV/optical SED shape of AGN accretion disk}
\label{sec:flex_disk}
\subsection{Motivation}
{\sc x-cigale} adopts a single fixed SED shape of an AGN accretion disk from \cite{schartmann05}, which is a broken power-law.
Although the \cite{schartmann05} recipe is a good approximation for the overall disk SED shape, it might not be sufficiently accurate for individual sources.
This is because the observed UV/optical slopes of type~1 quasars have non-negligible intrinsic dispersions \citep[e.g.,][]{elvis94}, possibly due to different black-hole masses, accretion rates, and spins \citep[e.g.,][]{koratkar99}.

\subsection{Sample and preliminary fitting}
\label{sec:quasar_sample}

\begin{table*}
\centering
\caption{Model parameters for the SDSS quasars}
\label{tab:par_quasar}
\begin{tabular}{llll} \hline\hline
Module & Parameter & Symbol & Values \\
\hline
    \multirow{6}{*}{\shortstack[l]{AGN (UV-to-IR) \\ SKIRTOR }}
    & AGN contribution to IR luminosity & $\fracA$ & 0.9999  \\
    & Viewing angle & $\theta$ & 30$^\circ$ \\
    & \multirow{2}{*}{\shortstack[l]{Polar-dust color excess}} & \multirow{2}{*}{\shortstack[l]{$E(B-V)_{\rm PD}$}} & \multirow{2}{*}{\shortstack[l]{0., 0.01, 0.02, 0.05, 0.1, 0.15, \\ 0.2, 0.3, 0.4, 0.5, 0.6 mag}} \\\\
    & \textbf{Intrinsic disk type} & -- & \textbf{\cite{schartmann05}} \\
    & \textbf{Deviation from the default UV/optical slope} 
    & $\boldsymbol{\delta_{\rm AGN}}$ & $\boldsymbol{-1}$\textbf{ to 1 (step 0.1)} \\
\hline
    \multirow{2}{*}{\shortstack[l]{X-ray}}
    & AGN photon index & $\Gamma$ & 1.8 \\
    & Maximum deviation from the $\ox$-$\luvr$ relation & $|\Delta \ox|_{\rm max}$ & 0.2 \\
\hline 
\end{tabular}
\begin{flushleft}
{\sc Note.} --- For parameters not listed here, we use the default values.
\textbf{Bold font} indicates new parameters in \cigv\ introduced in this work.
\end{flushleft}
\end{table*}

\begin{figure*}
    \centering
	\includegraphics[width=2\columnwidth]{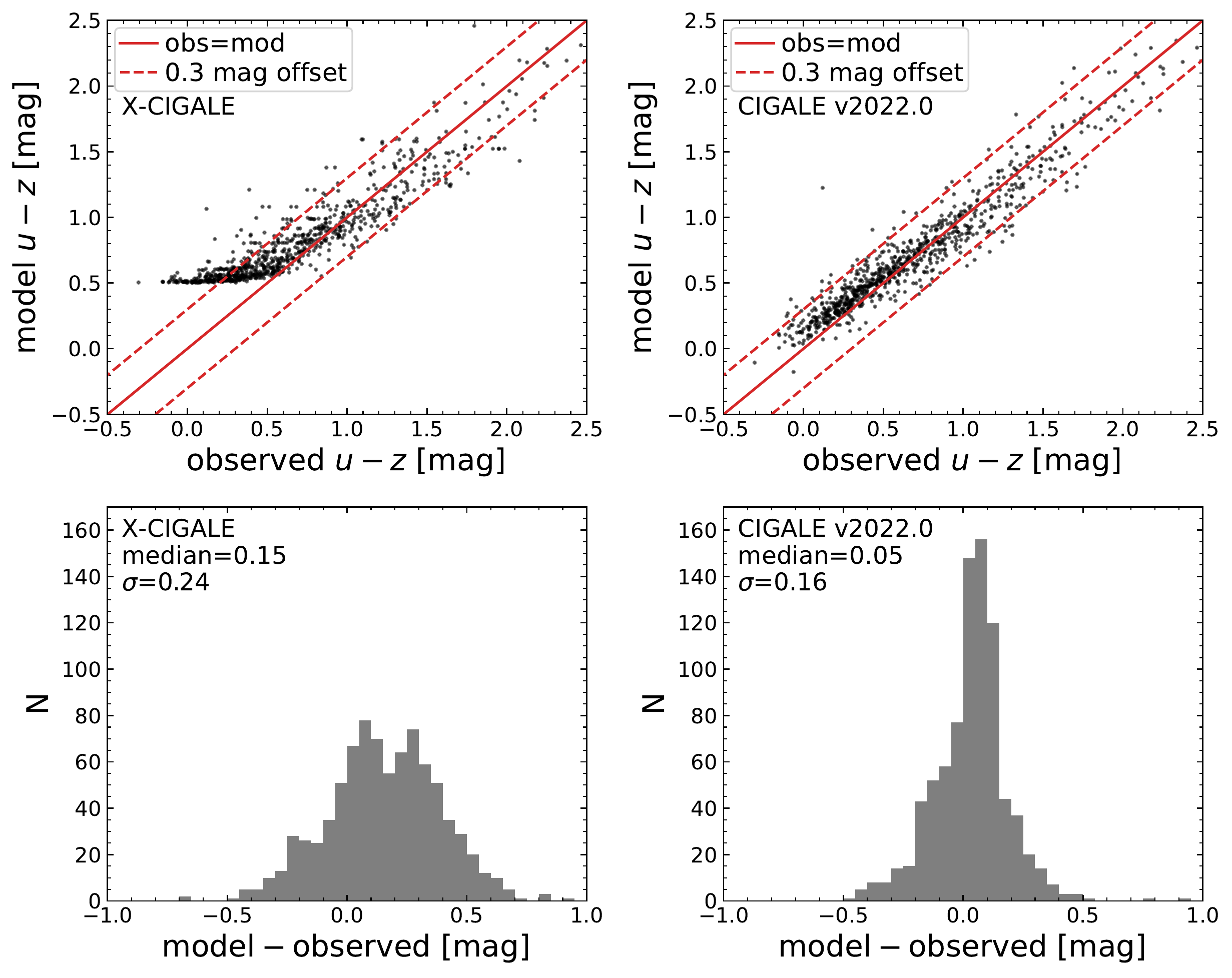}
    \caption{\textit{Top}: Model versus\ observed $u-z$ color of our SDSS quasars from \xcig\ (left) and \cigv\ (right).
    The red solid line indicates a model$=$observed relation; the red dashed lines indicate 0.3~mag offsets from this relation.
    We only plot points for sources having both $u$ and $z$ S/N$>5$.
    \textit{Bottom}: The distributions of model$-$observed $u-z$ color from \xcig\ (left) and \cigv\ (right).
    The median and standard-deviation values are labeled on each panel.
    The observed $u-z$ color can be bluer than $0.5$~mag. 
    This cannot be re-produced with \xcig, where the AGN disk SED shape is fixed. 
    But \cigv\ can account for these blue SEDs with $u-z<0.5$~mag.
    }
    \label{fig:quasar_obs_vs_mod}
\end{figure*}

\begin{figure*}
    \centering
	\includegraphics[width=\columnwidth]{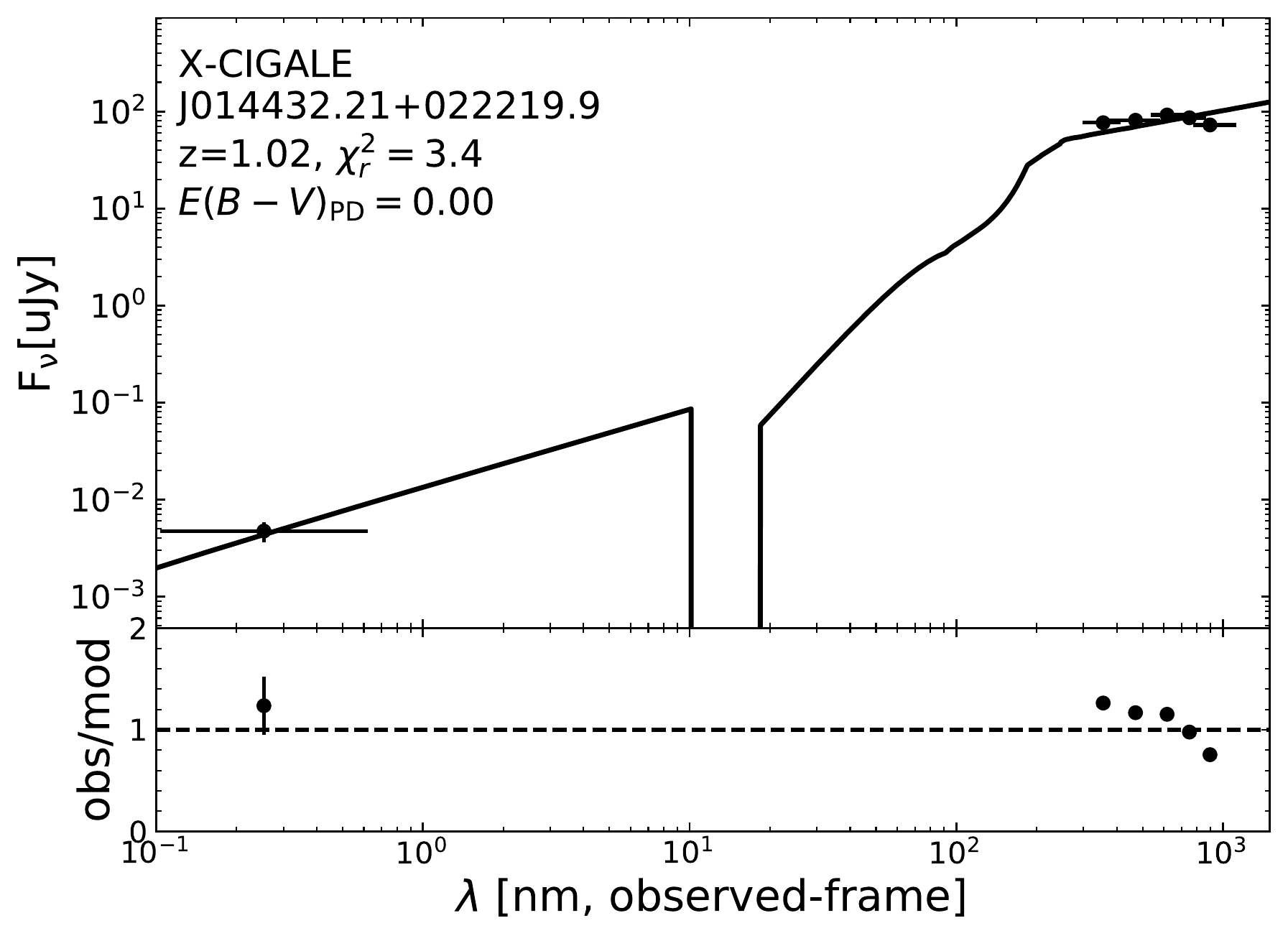}
	\includegraphics[width=\columnwidth]{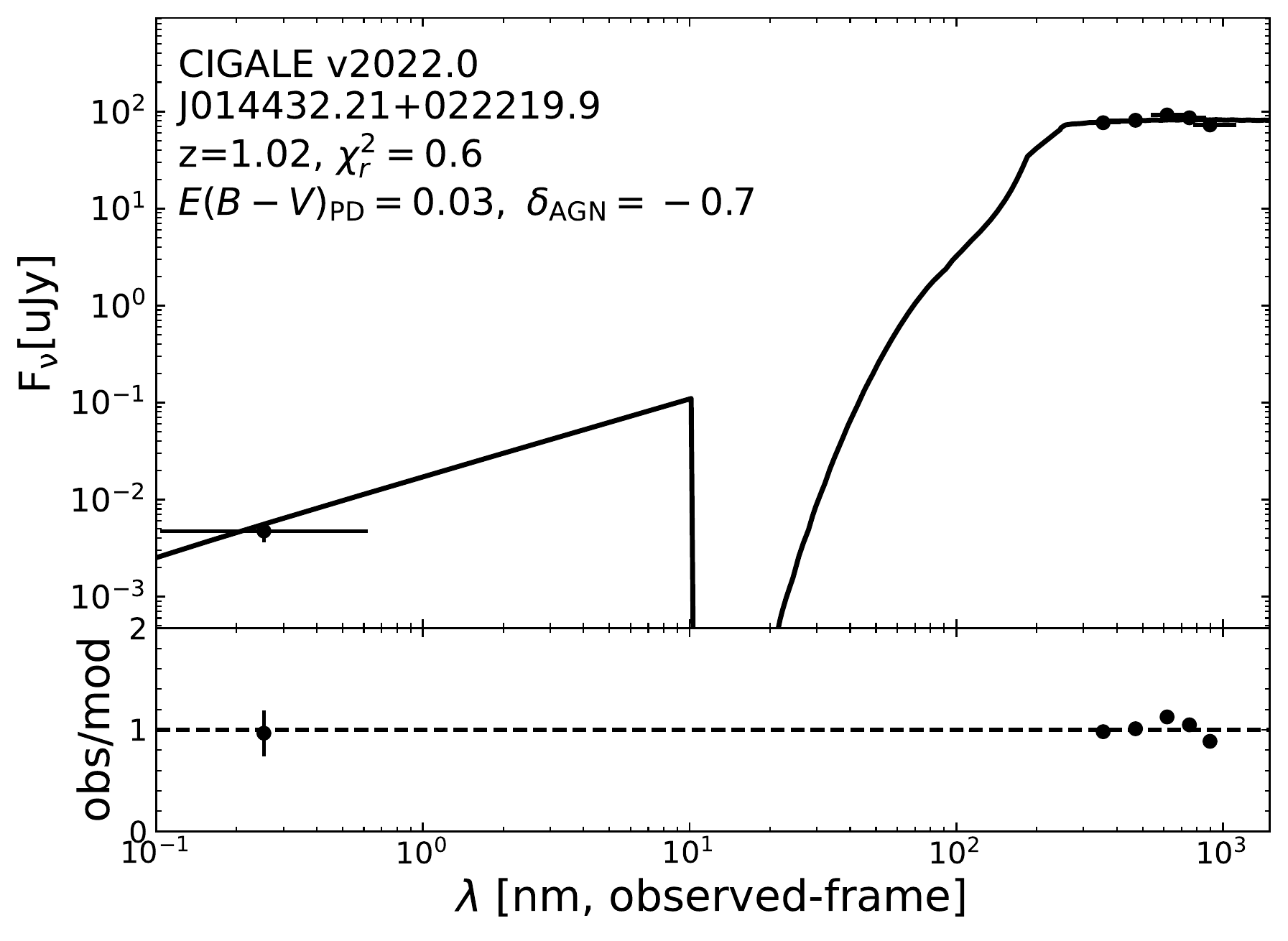}
    \caption{Example SED fits for an SDSS quasar from \xcig\ (left) and \cigv\ (right).
    This source has observed UV/optical SED bluer than 
    the model SED of \cite{schartmann05}, and thus it cannot be fitted well with \xcig. 
    \cigv\ allows a flexible model that can account for the blue SED shape.
    }
    \label{fig:quasar_sed}
\end{figure*}

To test whether this single spectral shape is sufficient to account for the observed SEDs, we use the SDSS DR14 type~1 quasar sample \citep{paris18} that has \xmm\ \xray\ detections (see \S3.1.1 of \citealt{yang20} for the details).
We further apply a magnitude cut ($r_{\rm AB}<21.8$~mag) and a redshift cut ($z>1$) to ensure that the observed SEDs are dominated by AGNs, as SDSS normal galaxies with $r_{\rm AB}<21.8$~mag are always below $z=1$ \citep[e.g.,][]{sheldon12}.
\scd{Therefore, these cuts allow us to model SEDs with pure AGN templates as below, avoiding potential degeneracy issues (see \S\ref{sec:res_agn_sed}).}

The final sample has 1080 sources. 
We run \xcig\ on the SDSS $ugriz$ and the 2--10~keV fluxes. 
The inclusion of \xray\ photometry is to better constrain the AGN intrinsic emission.
The fitting parameters are listed in Table~\ref{tab:par_quasar}.
We set $\fracA$ (fractional AGN IR luminosity) to a value close to unity (0.9999), so that the observed UV/optical SED is totally AGN dominated, which is the case for the SDSS quasars \fst{after our magnitude and redshift cuts}. 
We also allow different levels of polar-dust extinction. 

We compare the resulting \xcig\ model versus\ observed $u-z$ colors in Fig.~\ref{fig:quasar_obs_vs_mod} (left). 
In this figure, we only consider sources having both $u$ and $z$ signal-to-noise ratios (S/N) above 5.
One notable issue is that a ``plateau'' exists at model $u-z\approx 0.5$.
This is because the model $u-z$ cannot be bluer than the intrinsic disk color ($u-z=0.5$) but a significant fraction (32\%) of sources have observed $u-z<0.5$. 
Due to this issue, the offsets between the model and observed $u-z$ have a positive median value of 0.15.
Fig.~\ref{fig:quasar_sed} (left) shows an example SED. 
The observed SED is even bluer than the zero-extinction model SED. 
To address this issue, it is necessary to allow a flexible SED shape for AGN intrinsic-disk emission.

\subsection{Code improvement}
\label{sec:agn_sed_code}
To allow for deviations from the default \cite{schartmann05} optical spectral slope as suggested by the observed quasars (\S\ref{sec:quasar_sample}), we introduce a free parameter, $\delta_{\rm AGN}$, i.e.,
\begin{equation}
\label{eq:schartmann}
\lambda L_{\lambda} \propto\left\{\begin{array}{lr}
\lambda^{2} & 0.008 \leq \lambda \leq 0.05\ [\mu \mathrm{m}] \\
\lambda^{0.8} & 0.05<\lambda \leq 0.125\ [\mu \mathrm{m}] \\
\lambda^{-0.5+\delta_{\rm AGN}} & 0.125<\lambda \leq 10\ [\mu \mathrm{m}] \\
\lambda^{-3} & 10<\lambda \leq 1000\ [\mu \mathrm{m}]
\end{array}\right.
\end{equation}

In \cigv, we also allow the user to choose the disk continuum from the SKIRTOR model \citep{stalevski12, stalevski16}, i.e.,
\begin{equation}
\label{eq:skirtor}
\lambda L_{\lambda} \propto\left\{\begin{array}{lr}
\lambda^{1.2} & 0.008 \leq \lambda \leq 0.01\ [\mu \mathrm{m}] \\
\lambda^{0} & 0.01<\lambda \leq 0.1\ [\mu \mathrm{m}] \\
\lambda^{-0.5+\delta_{\rm AGN}} & 0.1<\lambda \leq 5\ [\mu \mathrm{m}] \\
\lambda^{-3} & 5<\lambda \leq 1000\ [\mu \mathrm{m}]
\end{array}\right.
\end{equation}
From Eqs.~\ref{eq:schartmann} and \ref{eq:skirtor} the major differences between the \cite{schartmann05} and SKIRTOR disk continuum are the wavelength boundaries and power-law indices at far-UV ($\lambda<125$~nm), where observational constraints are weaker compared to those at longer wavelengths \citep[e.g.,][]{stevans14, lusso15}. 

\subsection{Results and interpretation}
\label{sec:res_agn_sed}
\begin{figure}
    \centering
	\includegraphics[width=\columnwidth]{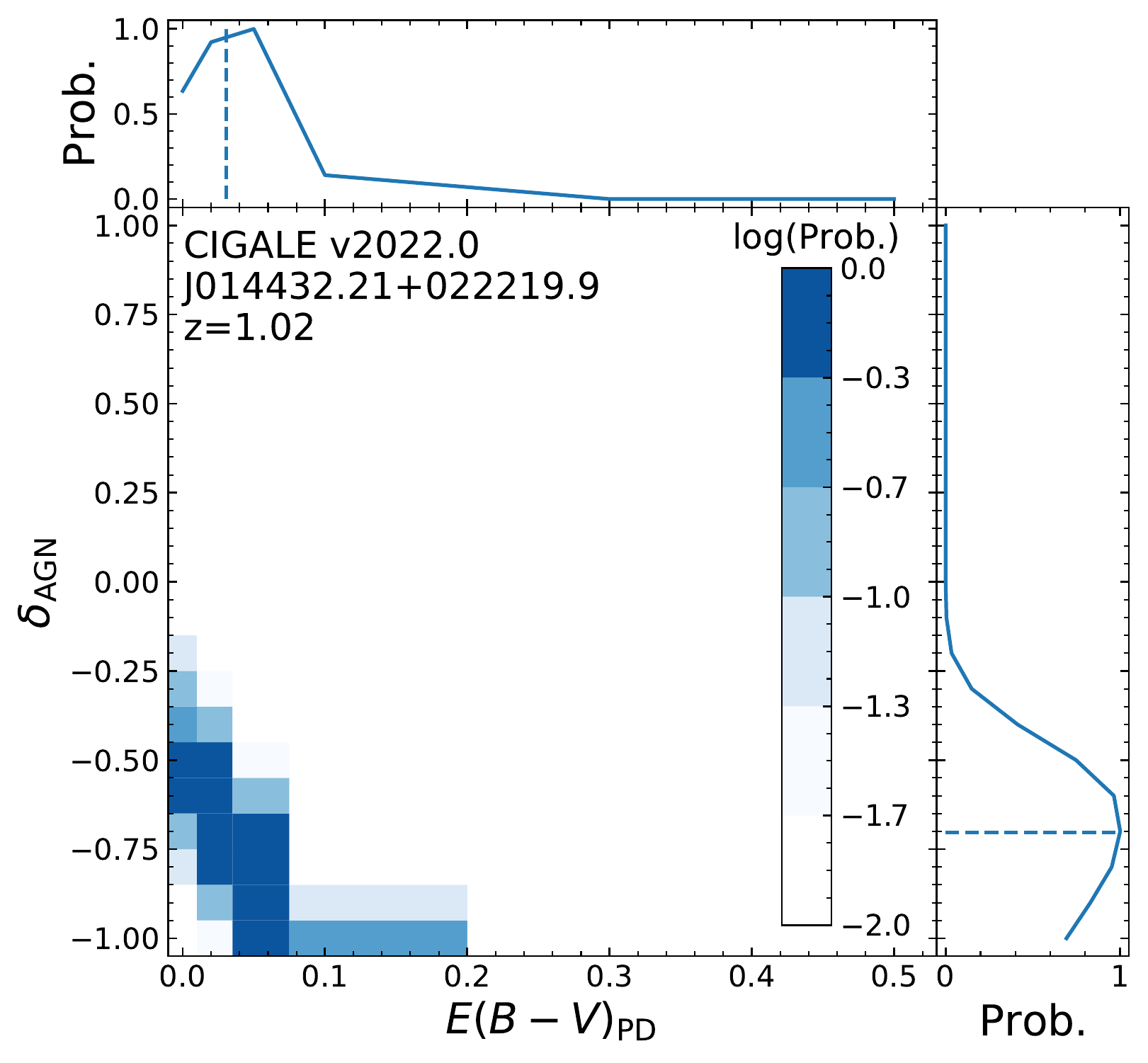}
    \caption{2D and 1D probability distributions of $\delta_{\rm AGN}$ and $E(B-V)_{\rm PD}$ for the SDSS quasar in Fig.~\ref{fig:quasar_sed}, generated from the \cigv\ run.
    The dashed line indicates the Bayesian (probability-weighted average) values of the two parameters. 
    From the color-coded 2D probability distribution, the $\delta_{\rm AGN}$ and $E(B-V)_{\rm PD}$ are anti-correlated, indicating a degeneracy between these two parameters.
    }
    \label{fig:quasar_2d_prob}
\end{figure}

\begin{figure}
    \centering
	\includegraphics[width=\columnwidth]{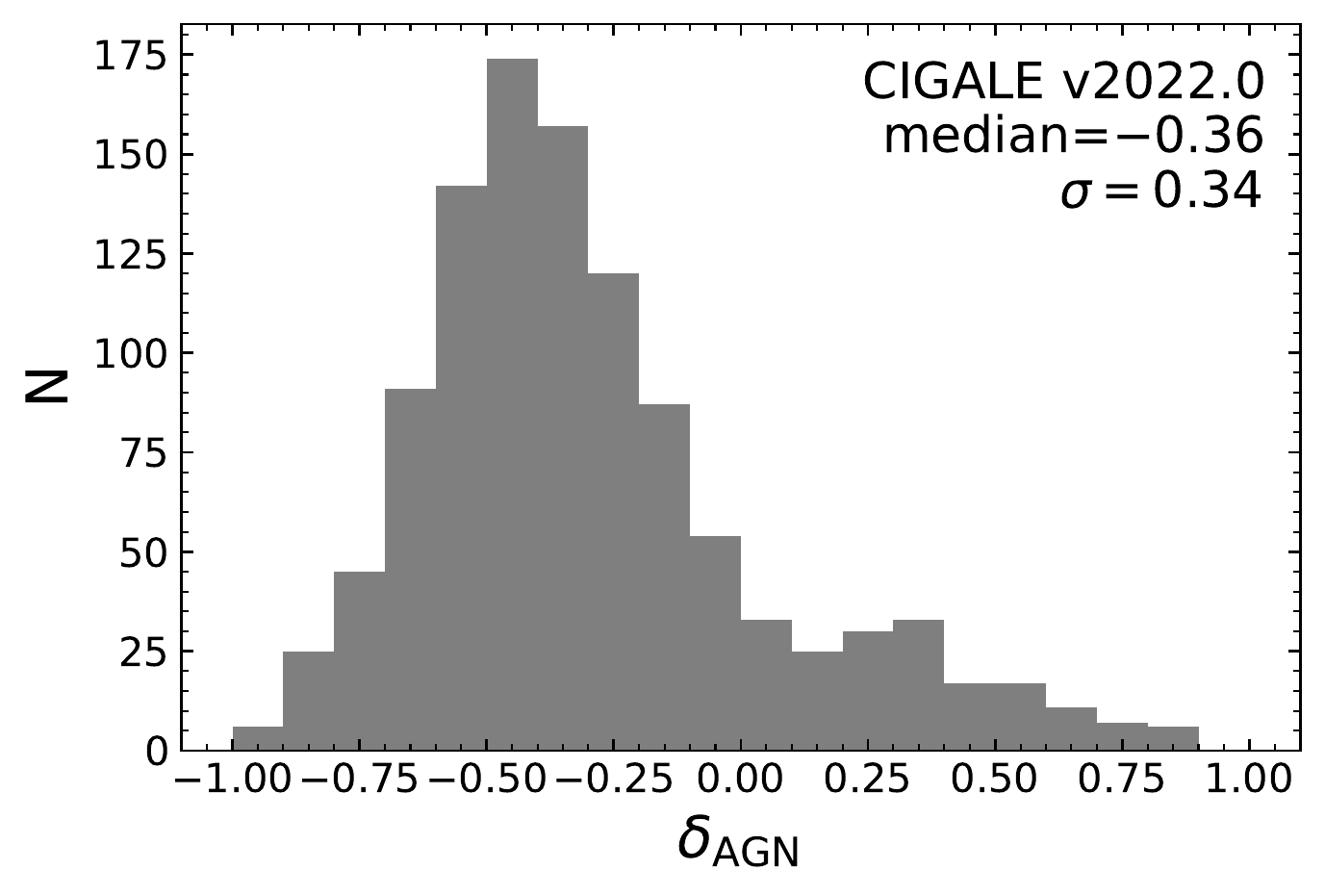}
	\includegraphics[width=\columnwidth]{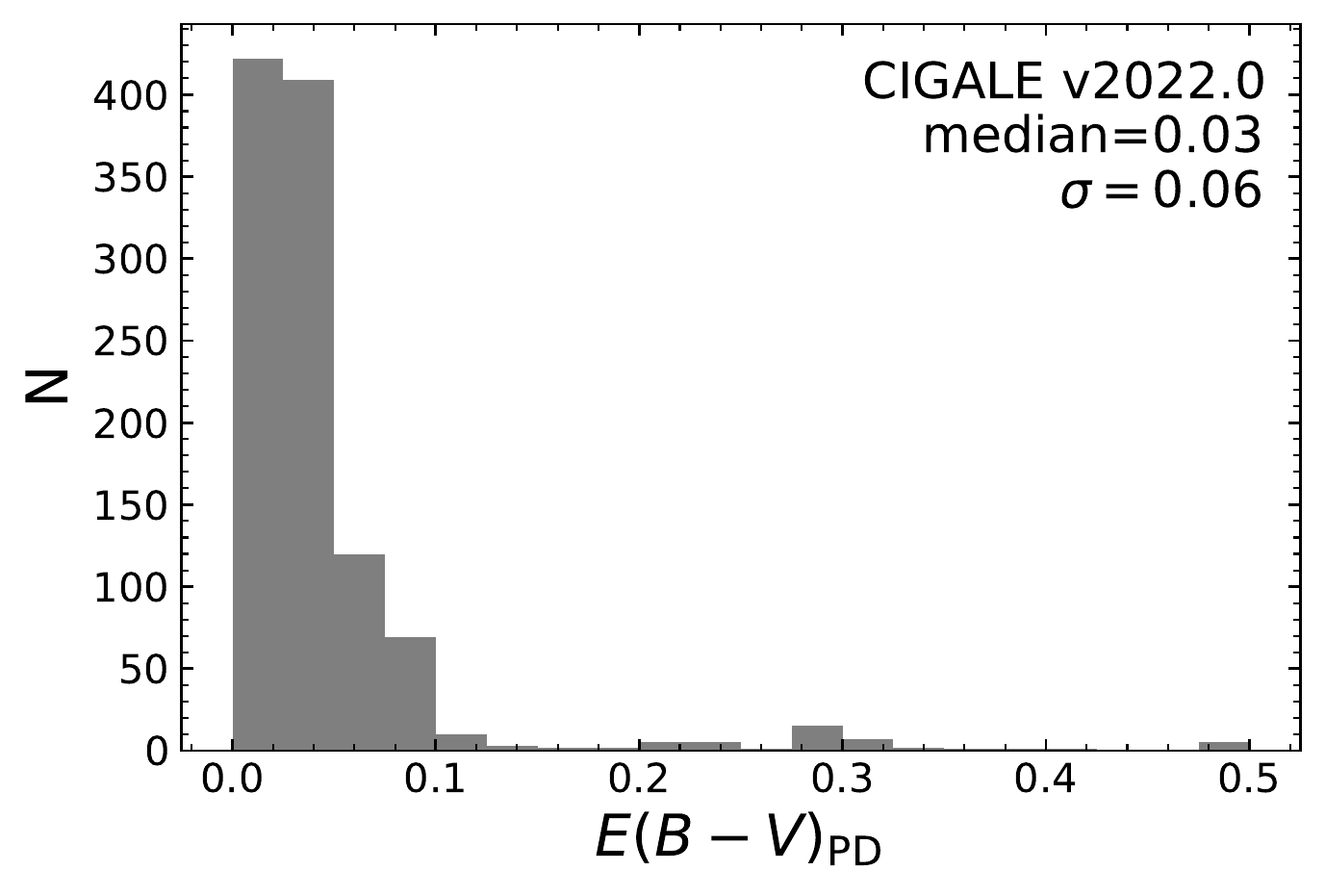}
    \caption{Distribution of the $\delta_{\rm AGN}$ and $E(B-V)_{\rm PD}$ for the SDSS quasar sample from the fits with \cigv. 
    The median and standard deviation values for the distributions are labeled on corresponding panels. 
    The $\delta_{\rm AGN}$ values tend to be negative (i.e., bluer than the default \citealt{schartmann05} SED model).
    }
    \label{fig:quasar_det_hist}
\end{figure}

We re-fit the SDSS quasar SEDs using \cigv. 
We still adopt the \cite{schartmann05} disk SED shape \fst{(Eq.~\ref{eq:schartmann})}, since it agrees better with observations \fst{than the SKIRTOR disk model} \citep[\fst{Eq.~\ref{eq:skirtor}}; e.g.,][]{duras17}.
We allow $\delta_{\rm AGN}$ to vary from $-1$ to 1 with a step of 0.1 (see Table~\ref{tab:par_quasar}).
This $\delta_{\rm AGN}$ range covers nearly all of the observed SED slope variations in the SDSS quasar sample of \citet[][see their Fig.~2]{davis07}. 

We compare the model versus\ observed $u-z$ colors in 
Fig.~\ref{fig:quasar_obs_vs_mod} (right). 
The model colors agree better with the observed colors than those from the \xcig\ fits.
The median offset of the model and observed $u-z$ is close to zero (\fst{0.05}), while the \xcig\ fits have a median of \fst{0.15}.
Also, the scatter has been reduced from \fst{0.24} (\xcig\ fits) to \fst{0.16} (\cigv\ fits).
Fig.~\ref{fig:quasar_sed} (right) shows an example SED fit. 
The observed SED can be well fitted with a reduced $\chi_r^2=0.6$ (the value is \fst{3.4} for the \xcig\ fit).
Fig.~\ref{fig:quasar_2d_prob} displays the 2D and 1D probability distributions of $\delta_{\rm AGN}$ and $E(B-V)_{\rm PD}$ of this example fit. 
There is an anti-correlation between $\delta_{\rm AGN}$ and $E(B-V)_{\rm PD}$ in the 2D probability distribution. 
This anti-correlation is expected, because, e.g., a higher $\delta_{\rm AGN}$ (redder intrinsic slope) and a lower $E(B-V)_{\rm PD}$ (weaker polar-dust extinction) can roughly cancel out the effect of each other. 
Therefore, there is a natural degeneracy between these two parameters. 

Fig.~\ref{fig:quasar_det_hist} displays the distribution of the fitted $\delta_{\rm AGN}$ and $E(B-V)_{\rm PD}$. 
The median of the $\delta_{\rm AGN}$ distribution is \fst{$-0.36$} with a significant scatter of \fst{0.34}. 
This negative median value could be intrinsic, as the default \cite{schartmann05} spectral shape is based on the observed quasar SEDs, without considering polar-dust extinction. 
Indeed, the polar-dust extinctions are non-negligible [median $E(B-V)_{\rm PD}=0.03$], although heavy extinctions of $E(B-V)_{\rm PD}>0.1$ are rare (\fst{6\%}). 
However, we note that the degeneracy between $\delta_{\rm AGN}$ and $E(B-V)_{\rm PD}$ (e.g., Fig.~\ref{fig:quasar_2d_prob}) might also contribute to the negative trend of $\delta_{\rm AGN}$.
We caution that both of the $\delta_{\rm AGN}$ and $E(B-V)_{\rm PD}$ distributions quantitatively depend on the adopted specific extinction law. 
We adopt the default SMC law here, and refer to \cite{buat21} for a detailed discussion of the effects of different laws. 

Given the degeneracy between $\delta_{\rm AGN}$ and $E(B-V)_{\rm PD}$, one might think of adapting the original disk SED shapes (Schartmann and SKIRTOR; \S\ref{sec:agn_sed_code}) to the bluest variation in our sample and attributing all observed SED variations to $E(B-V)_{\rm PD}$. 
This idea has the benefit of simplicity, but we do not adopt it for three reasons. 
First, it is unlikely that all AGNs share the same intrinsic SED shape over wide ranges of BH masses and accretion rates \citep[e.g.,][]{whiting01, richards03}.
Second, the current approach is more flexible than adapting the template, and the philosophy of \cig\ highlights flexibility rather than simplicity. 
Third, although $E(B-V)_{\rm PD}$ and $\delta_{\rm AGN}$ are degenerate over UV/optical wavelengths, they can be better differentiated given excellent IR coverage, because dust-reddened AGNs have polar-dust IR re-emission (included in \cig) but intrinsically red AGNs do not. 
\scd{In \cig, polar dust is an obscuration structure with optical depth much smaller than the torus (see \S2.4 of \citealt{yang20} for details). 
\cig\ assumes that the polar-dust IR emission follows a ``grey body'' model \citep[e.g.,][]{casey12} with temperature and emissivity as free parameters. 
}
Future IR missions, e.g., \jwst\ and \textit{Origins}, will be able to detect (tightly constrain) the polar-dust IR re-emission and thereby differentiate the two cases of dust-reddened vs.\ intrinsically red SEDs.

We set the default $\delta_{\rm AGN}$ and $E(B-V)_{\rm PD}$ to the median values of our fits, i.e., \fst{$\delta_{\rm AGN}=-0.36$ and $E(B-V)_{\rm PD}=0.03$}. 
When fitting type~1 AGNs, the user is recommended to adopt multiple values of these two parameters (such as those in Table~\ref{tab:par_quasar}), because both parameters have significant scatters based on our fits (see Fig.~\ref{fig:quasar_det_hist}).
When fitting type~2 AGNs, for which the AGN disk emission is almost entirely obscured by the dusty torus, the user can keep the default $\delta_{\rm AGN}$ to save memory and reduce computational time. 
However, it is still recommended to adopt multiple values of $E(B-V)_{\rm PD}$ when mid-to-far IR coverage is available for the type~2 sources, because $E(B-V)_{\rm PD}$ sets the strength of polar-dust re-emission that could contribute significantly at IR wavelengths. 

\begin{figure*}
    \centering
	\includegraphics[width=\columnwidth]{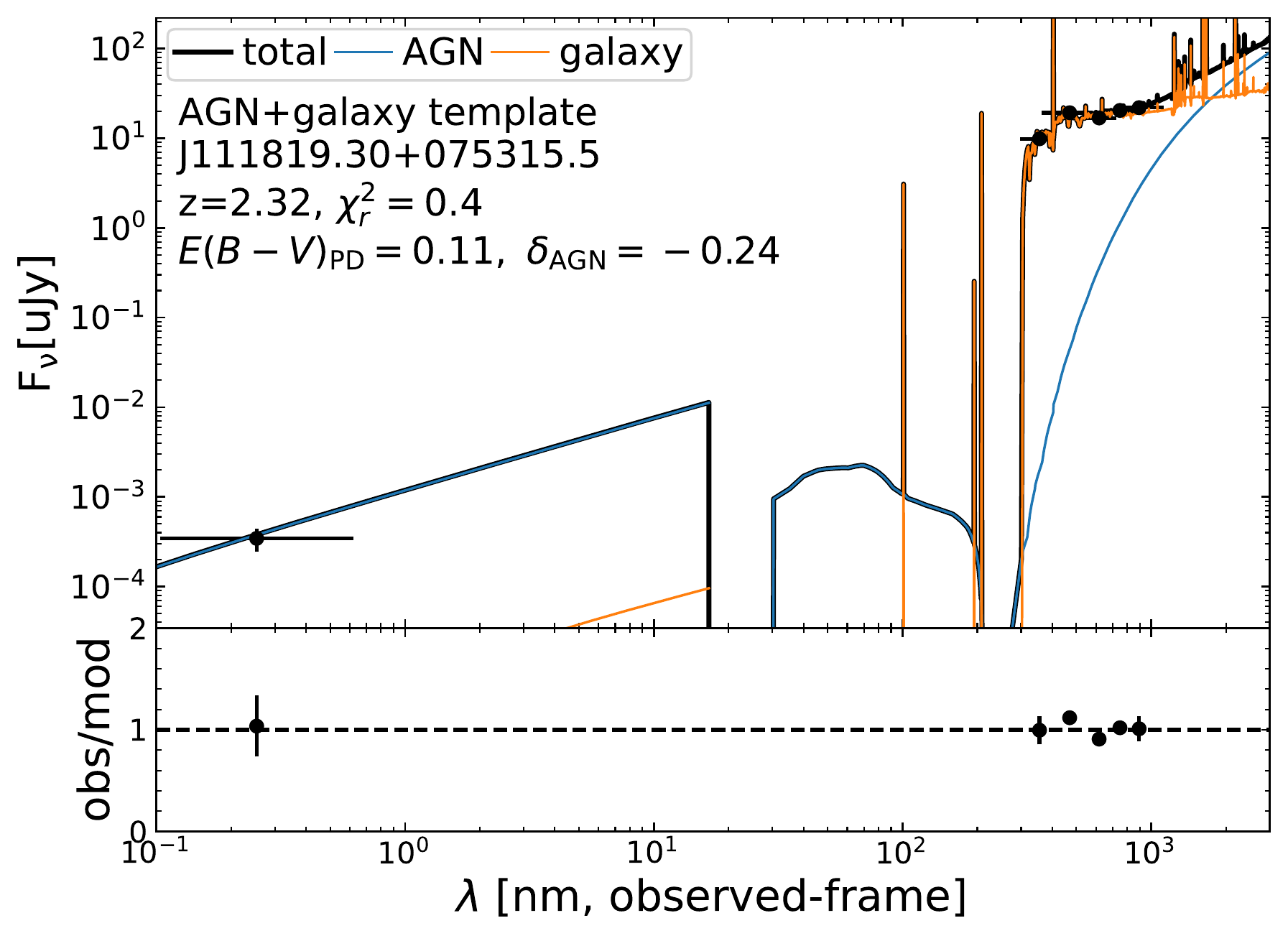}
	\includegraphics[width=\columnwidth]{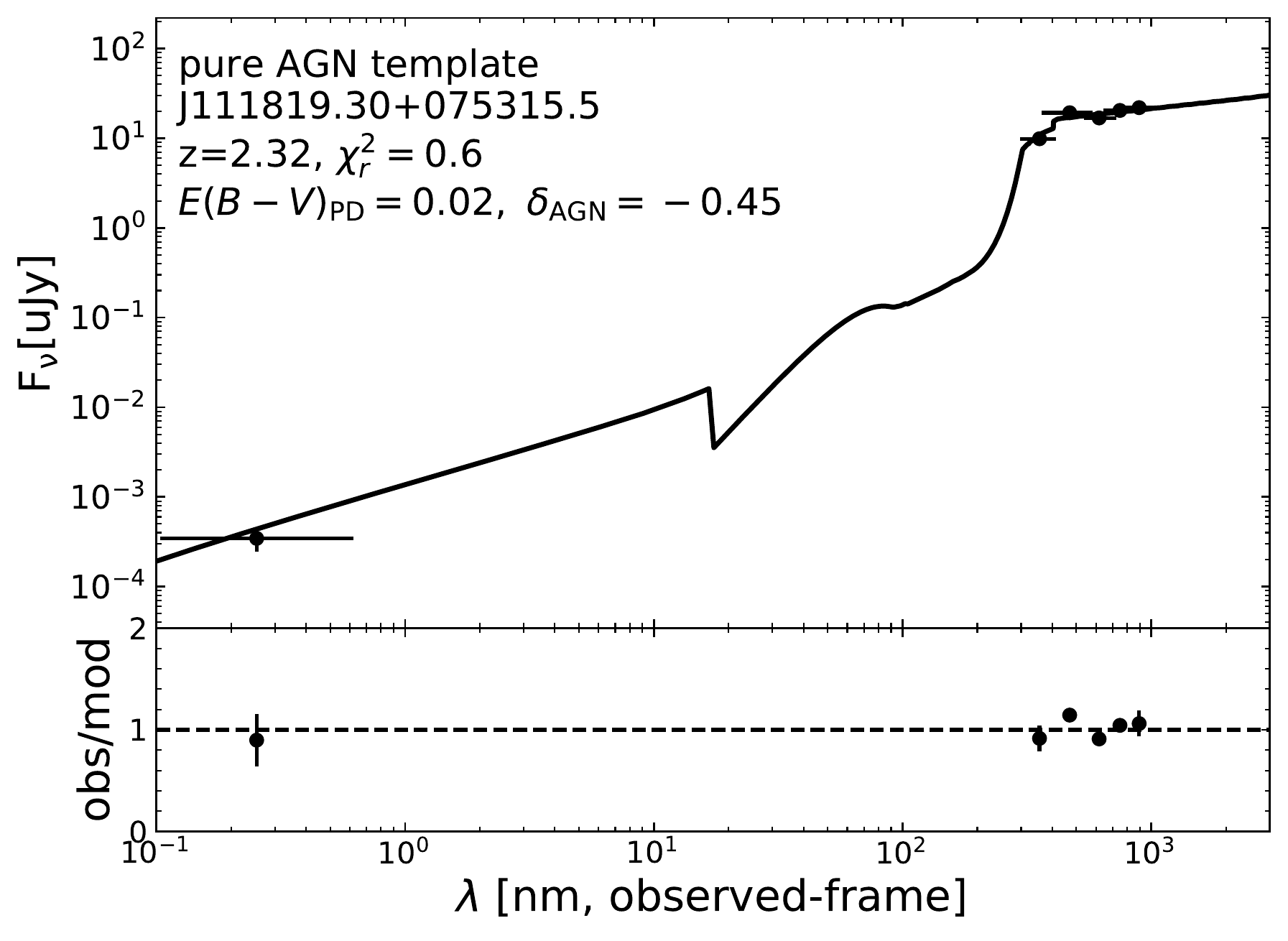}
    \caption{\scd{Example SED fits for an SDSS quasar using AGN$+$galaxy mixed models (left) and pure AGN models (right; the approach in our analysis).
    Both fits can explain the observed SED data, indicating that quasar and blue star-forming templates are degenerate.  
    The resulting $E(B-V)_{\rm PD}$ and $\delta_{\rm AGN}$ (as labeled) are quite different for the two fits, indicating the degeneracy can significantly affect these parameters.
    Since the galaxy component on the left panel is unrealistic (see \S\ref{sec:quasar_sample}), we adopt pure AGN models in our analysis to avoid the degeneracy issue.    
    }
    }
    \label{fig:quasar_sed_dg}
\end{figure*}

Finally, we stress the importance of our redshift and magnitude cuts (\S\ref{sec:quasar_sample}).
These cuts guarantee the observed SEDs are dominated by AGNs, allowing us to model the data with pure AGN templates. 
Actually, we have tested fits with AGN$+$galaxy mixed templates by freeing $\fracA$, and found our parameters of interest ($\delta_{\rm AGN}$ and $E(B-V)_{\rm PD}$) could be significantly affected. 
This is because a blue observed SED can be either explained by a low-dust star-forming galaxy component or an AGN component. 
Fig.~\ref{fig:quasar_sed_dg} displays such an AGN-galaxy degeneracy for an SDSS quasar, for which $\delta_{\rm AGN}$ and $E(B-V)_{\rm PD}$ are strongly affected when a galaxy component is allowed. 
Therefore, our redshift/magnitude cuts and pure-AGN approach are crucial for our investigation of $\delta_{\rm AGN}$ and $E(B-V)_{\rm PD}$. 
The user who is interested in these parameters should be cautious of the degeneracy effects, when modeling SEDs that have non-negligible galaxy components.
On the other hand, some studies indicate that some other source properties such as $\fracA$, AGN bolometric luminosity ($L_{\rm AGN}$), and SFR are not strongly affected by the degeneracy issue, especially when good multiwavelength coverage is available \citep[e.g.,][]{mountrichas21b, yang21, thorne22}. 
This is understandable, considering that those properties can be constrained by multiwavelength data simultaneously, e.g., $L_{\rm AGN}$ is related to \xray, UV/optical, and IR wavelengths. 
However, in contrast, $\delta_{\rm AGN}$ and $E(B-V)_{\rm PD}$ are very sensitive to the detailed SED-shape modeling at UV/optical wavelengths, and thus they are more strongly affected by the AGN-galaxy degeneracy than properties like $L_{\rm AGN}$.

\section{AGN radio emission}
\label{sec:agn_radio}

\subsection{Motivation}
\label{sec:radio_mot}
{\sc x-cigale} can only account for radio emission from star formation \citep[SF;][]{boquien19}. 
This SF radio emission has two components:
one is a thermal component contributed by the ``nebular'' module;
the other is a synchrotron component contributed by the ``radio'' module. 
The latter is often dominant, and is calculated in \xcig\ using the radio-IR correlation parameter $q_{\rm IR}$ \citep[e.g.,][]{helou85}, i.e., 
\begin{equation}
\label{eq:qir}
    q_{\rm IR} = \log \left( \frac{L_{\rm SF, IR}}
                         {L_{\nu, 21\rm cm}\times 3.75\times 10^{12}\rm\ Hz} 
             \right)
\end{equation}
where $L_{\rm SF, IR}$ is the total star-forming IR luminosity (mostly in FIR) and $L_{\nu, 21\rm cm}$ is the corresponding radio synchrotron luminosity at 21~cm (1.4~GHz).
The default value of $q_{\rm IR}$ is 2.58 in \xcig. 
Besides $q_{\rm IR}$, which sets the normalization at 21~cm, there is another free
parameter ($\alpha_{\rm SF}$) that controls the power-law slope of the SF 
synchrotron emission, i.e., 
\begin{equation}
    L_{\nu, \rm SF} \propto \nu^{-\alpha_{\rm SF}}.
\end{equation}
The default value of $\alpha_{\rm SF}$ is 0.8 in both \xcig\ and \cigv.

{\sc x-cigale} does not have AGN emission at radio wavelengths. 
However, AGNs may have powerful jets that emit strong radio radiation (i.e., radio-loud AGNs), and the jets can play an important role in AGN-galaxy coevolution \citep[e.g.,][]{fabian12}.
The physical origin of AGN radio jets is still controversial. 
One popular theory is the BZ (Blandford-Znajek) process \citep[e.g.,][]{blandford77, blandford19}. 
The BZ mechanism considers that the jet is powered by the rotational energy of the black hole (BH) through the magnetic field threading the horizon \citep[e.g.,][]{davis20}.
Recent observations suggest that the magnetic flux/topology close to the BH instead of the BH spin could be the determining factor of the jet-launching process \citep[e.g.,][]{zhu20}.
Besides the jets, other processes such as AGN winds, coronae, and shocks can also emit at radio wavelengths \citep[e.g.,][]{panessa19}.  

\subsection{Sample and preliminary fitting}
\label{sec:radio_sample}

\begin{table*}
\centering
\caption{Model parameters for the COSMOS radio sources}
\label{tab:par_radio}
\begin{tabular}{llll} \hline\hline
Module & Parameter & Symbol & Values \\
\hline
\multirow{2}{*}{\shortstack[l]{Star formation history\\
                               $\mathrm{SFR}\propto t \exp(-t/\tau)$ }}
    & Stellar e-folding time & $\tau_{\rm star}$ & 0.1, 0.5, 1, 5 Gyr\\
    & Stellar age & $t_{\rm star}$  
            & 0.5, 1, 3, 5, 7 Gyr\\ 
\hline
\multirow{2}{*}{\shortstack[l]{Simple stellar population\\ 
    \cite{bruzual03}}}
    & Initial mass function & $-$ & \cite{chabrier03} \\
    & Metallicity & $Z$ & 0.02 \\
\hline
\multirow{2}{*}{\shortstack[l]{Dust attenuation \\ 
                \cite{calzetti00} }}
    & \multirow{2}{*}{Color excess of the nebular lines} & 
        \multirow{2}{*}{$E(B-V)$} &
        \multirow{2}{*}{\shortstack[l]{0.05, 0.1, 0.2, \\
                                    0.3, 0.4, 0.5, 0.6 mag}} \\\\
\hline
\multirow{2}{*}{\shortstack[l]{Galactic dust emission: \\ \cite{dale14}}}
    & \multirow{2}{*}{\shortstack[l]{Slope in $dM_{\rm dust} \propto U^{-\alpha} dU$}}
    & \multirow{2}{*}{\shortstack[l]{$\alpha$}}
    & \multirow{2}{*}{\shortstack[l]{2}}
    \\\\
\hline
\multirow{3}{*}{\shortstack[l]{AGN (UV-to-IR) \\ SKIRTOR }}
    & AGN contribution to IR luminosity & $\fracA$ & 0--0.99 (step 0.1)  \\
    & Viewing angle & $\theta$ & 30$^\circ$, 70$^\circ$ \\
    & \multirow{1}{*}{\shortstack[l]{Polar-dust color excess}} & \multirow{1}{*}{\shortstack[l]{$E(B-V)_{\rm PD}$}} & \multirow{1}{*}{\shortstack[l]{0, 0.2, 0.4 mag}} \\
\hline
    \multirow{5}{*}{\shortstack[l]{Radio}}
        & SF radio-IR correlation parameter & $q_{\rm IR}$ & 2.4, 2.5, 2.6, 2.7 \\
        & SF power-law slope & $\alpha_{\rm SF}$ & 0.8 \\
        & \multirow{2}{*}{\shortstack[l]{\textbf{Radio-loudness parameter}}} & 
    \multirow{2}{*}{\shortstack[l]{$\boldsymbol{R_{\rm AGN}}$}} & 
    \multirow{2}{*}{\shortstack[l]{\textbf{0.01, 0.02, 0.05, 0.1, 0.2, 0.5,} 
        \\ \textbf{..., 1000, 2000, 5000, 10000}}} \\\\
        & \textbf{AGN power-law slope} & $\boldsymbol{\alpha_{\rm AGN}}$ & \textbf{0.7} \\
\hline 
\end{tabular}
\begin{flushleft}
{\sc Note.} --- For parameters not listed here, we use the default values.
\textbf{Bold font} indicates new parameters in \cigv\ introduced in this work.
\end{flushleft}
\end{table*}

\begin{figure*}
    \centering
	\includegraphics[width=2\columnwidth]{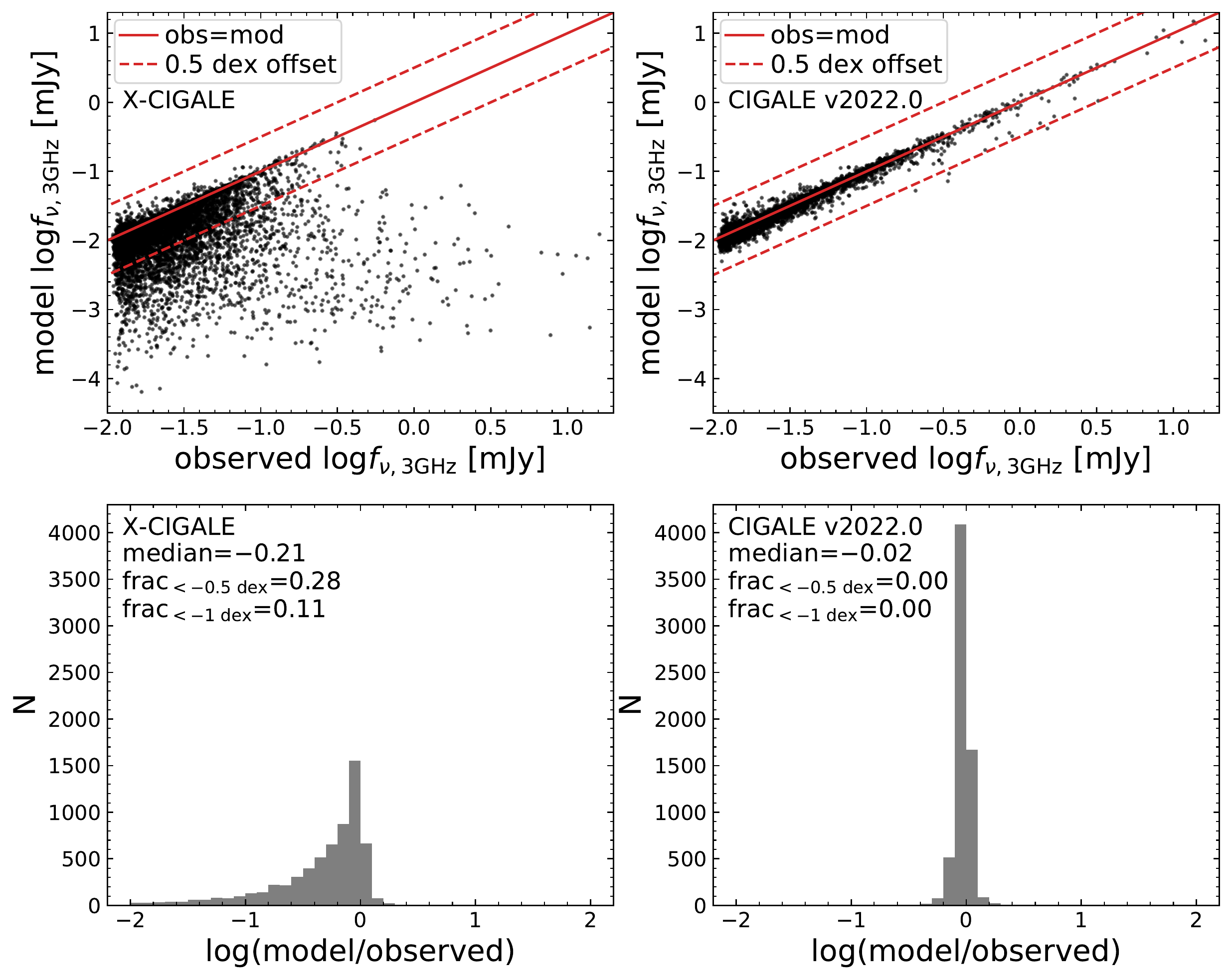}
    \caption{\textit{Top}: Model versus\ observed 3~GHz flux density of our COSMOS radio sources from \xcig\ (left) and \cigv\ (right).
    The red solid line indicates a model$=$observed relation; 
    the red dashed lines indicate 0.3~dex offsets from this relation.
    \textit{Bottom}: The distributions of logarithmic model/observed $f_{\nu, \rm 3GHz}$ from the \xcig\ (left) and \cigv\ (right).
    \xcig\ systematically underestimates $f_{\nu, \rm 3GHz}$,
    indicating the presence of radio AGN emission. 
    This underestimation does not exist for the fits of \cigv, which has an AGN radio component. 
    }
    \label{fig:radio_obs_vs_mod}
\end{figure*}

\begin{figure*}
    \centering
	\includegraphics[width=2\columnwidth]{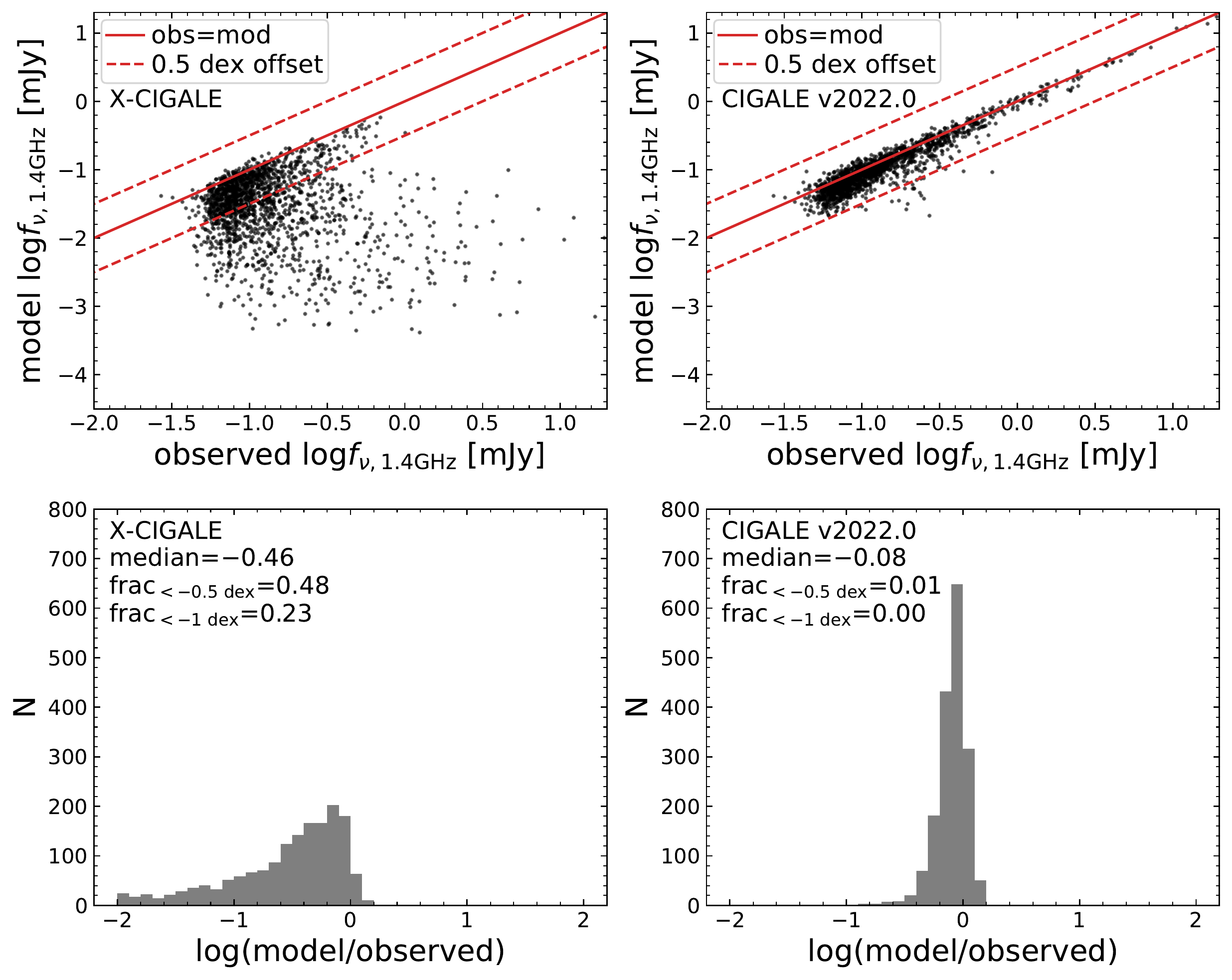}
    \caption{Same format as Fig.~\ref{fig:radio_obs_vs_mod} but for the 1.4~GHz band. 
    }
    \label{fig:radio_1p4G_obs_vs_mod}
\end{figure*}

\begin{figure*}
    \centering
	\includegraphics[width=\columnwidth]{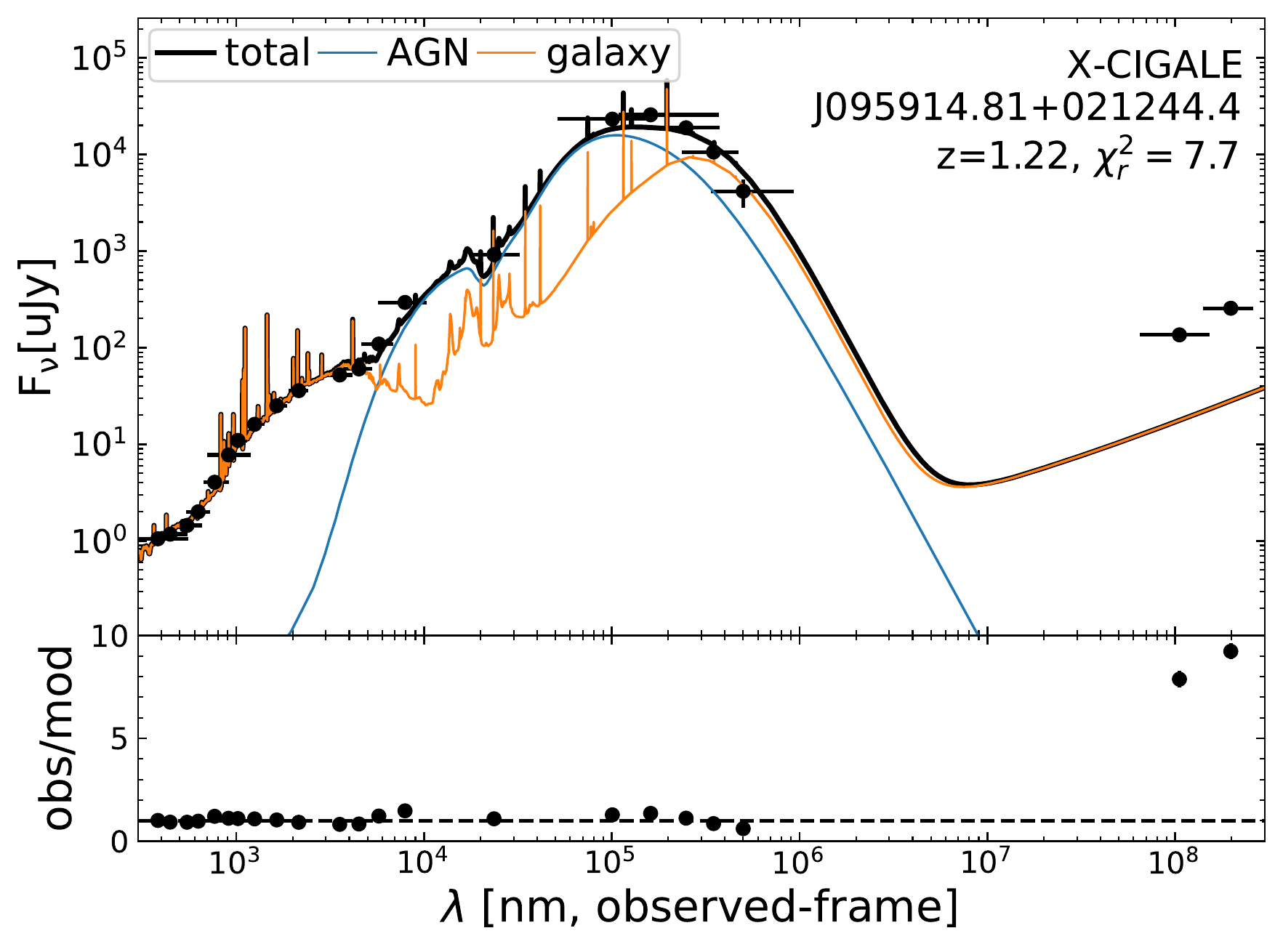}
	\includegraphics[width=\columnwidth]{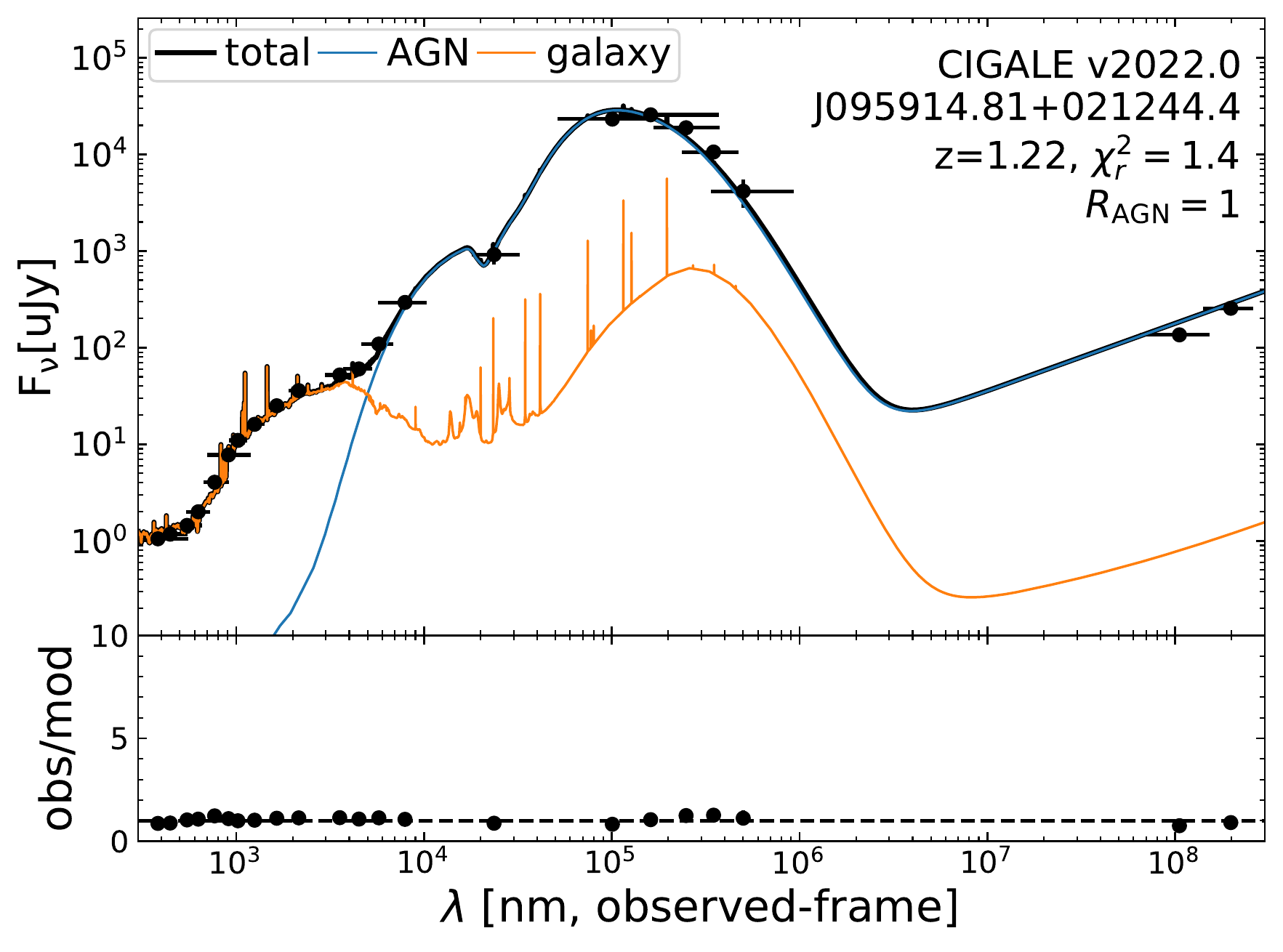}
    \caption{Example SED fits for a COSMOS radio source
    from \xcig\ (left) and \cigv\ (right).
    The observed radio fluxes are much higher than the model ones from \xcig.
    \cigv\ accounts for this radio excess with an AGN radio component.
    Also, compared to the \cigv\ fit, the \xcig\ fit has much stronger galaxy IR component, because the high observed radio fluxes force an elevated galaxy IR emission via $q_{\rm IR}$ (Eq.~\ref{eq:qir}).
    }
    \label{fig:radio_sed}
\end{figure*}

Similar to the procedures of the previous sections (\S\ref{sec:xray_ani}, \S\ref{sec:xrb}, and \S\ref{sec:flex_disk}), 
we first compile a proper radio-selected sample and then perform SED modeling with \xcig\ in this section. 
We adopt all of the $>5\sigma$ radio detections in the VLA-COSMOS 3~GHz Large Project \citep{smolcic17, smolcic17b}.
We also collect the VLA 1.4~GHz fluxes when available from \cite{schinnerer10}.
\cite{delvecchio17} matched the radio sources to the COSMOS2015 catalog \citep{laigle16}. 
We adopt these matching results and obtain the UV-to-IRAC4 broad-band photometry (14 bands) from COSMOS2015. 
We discard the radio sources without COSMOS2015 counterparts, as the UV-to-IRAC4 data are necessary to model the stellar population. 
The sample contains 6497 sources in total. 
We also include \spitzer/MIPS (24~$\mu$m), \herschel/PACS (100~$\mu$m and 160~$\mu$m), and \herschel/SPIRE (250~$\mu$m, 350~$\mu$m, and 500~$\mu$m) photometry from the ``super-deblended'' catalog of \cite{jin18}.
We do not include \xray\ fluxes here due to the reason presented in \S\ref{sec:res_radio}, 
i.e., we want to keep our SED fits and subsequent source classifications independent from the \xray\ information.
We adopt the redshift measurements from \cite{delvecchio17}, which are spec-$z$ (if available) or photo-$z$.
The median redshift is 1.18 and the 10\%--90\% percentile range is $z=0.42$--2.56. 

We first fit the photometric data above with \xcig.
The model parameters are listed in Table~\ref{tab:par_radio}.
We set $q_{\rm IR}$ to a range of 2.4--2.7 (step 0.1), based on the observations of \cite{delvecchio21}. 
Fig.~\ref{fig:radio_obs_vs_mod} (left) \fst{and Fig.~\ref{fig:radio_1p4G_obs_vs_mod} (left)} display the resulting  model fluxes versus the observed values \fst{for 3~GHz and 1.4~GHz, respectively.}
The model fluxes are systematically lower than the observed ones, e.g., 28\% (11\%) of sources have observed 3~GHz fluxes more than 3 (10) times higher than the model fluxes. 
In contrast, no sources have model fluxes $>3$ times higher than the observed values. 
This result of ``radio excess'' strongly indicates that an AGN radio component is needed to explain the observed radio fluxes for many sources. 
Fig.~\ref{fig:radio_sed} (left) shows an example SED fit with significant radio excess. 

\subsection{Code improvement}
\label{sec:radio_code}
We add a new AGN component to the radio module of \xcig. 
To quantitatively model AGN radio emission, we employ the radio-loudness 
parameter, $R$, defined as \citep[e.g.,][]{ballo12},
\begin{equation}
\label{eq:R_agn}
    R_{\rm AGN} = \frac{\lrr}{\luvr},
\end{equation}\\
where $\lrr$ and $\luvr$ are the monochromatic AGN luminosities 
per frequency at rest-frame 5~GHz and 2500~$\rm \AA$, respectively. 
$R_{\rm AGN}$ is a free parameter that allows any values $\geq 0$ 
($R_{\rm AGN}=0$ means no AGN radio emission). 

Here, we adopt $\luvr$ as the intrinsic (polar-dust absorption corrected) luminosity observed at a viewing angle of $30^\circ$, and this quantity is available for \xcig\ models (see \S\ref{sec:xray_ani}). 
This definition ensures that $R$ is a physical quantity inherent to the
AGN itself and does not depend on the viewing angle 
\citep[e.g.,][]{padovani16, padovani17}.
Therefore, $R$ works consistently for both type~1 and type~2 AGNs. 
Currently, we assume $\lrr$ is isotropic in \xcig. 
In the future, we will model the radio anisotropy, which can be important
for, e.g., blazars and BL~Lac objects. 

We assume a power-law AGN SED over the wavelength range of 0.1--1000~mm, i.e., 
\begin{equation}
\label{eq:alpha_agn}
    L_{\nu, \rm AGN} \propto \nu^{-\alpha_{\rm AGN}},
\end{equation}
where we allow the user to freely set the $\alpha_{\rm AGN}$ slope. 
We set the default value as $\alpha_{\rm AGN}=0.7$ \citep[e.g.,][]{randall12, tiwari19}. 
\fst{We caution that the power-law shape is an overall simplistic assumption, as the real AGN radio SEDs might be more complicated. 
The formula in Eq.~\ref{eq:alpha_agn} mainly serves as a correction for the AGN contribution to radio fluxes, especially for the cases where only one or two radio bands are available like our COSMOS radio sample. 
In the future, we will explore more realistic and complicated radio models based on multi-band radio data.
}

\subsection{Results and interpretation}
\label{sec:res_radio}

\begin{figure}
    \centering
	\includegraphics[width=\columnwidth]{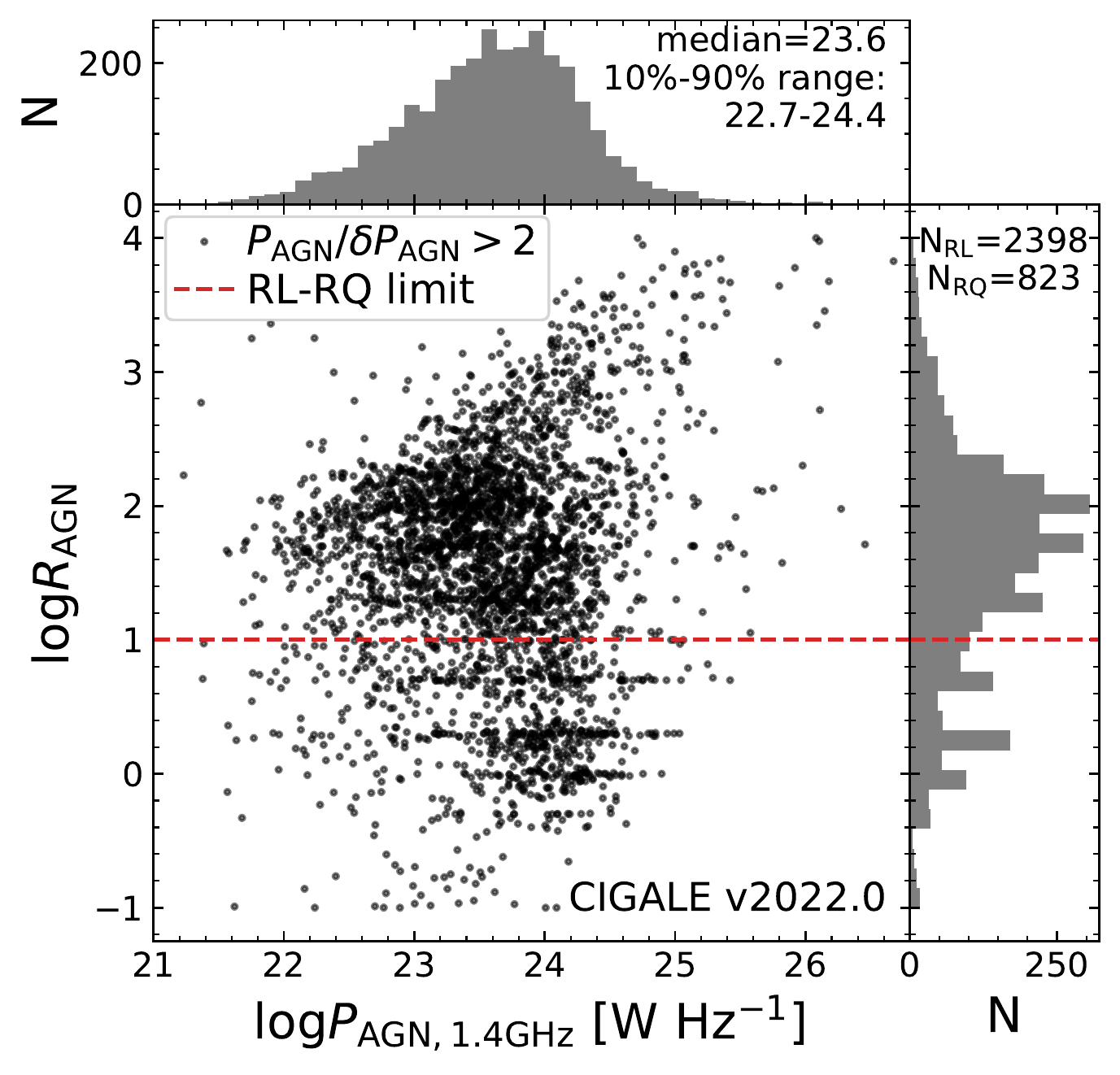}
    \caption{$R_{\rm AGN}$ versus AGN 1.4~GHz power and their distributions. 
    Only the sources with a significant AGN component ($\pagn/\delta \pagn>2$) are displayed. 
    The red dashed horizontal line indicates the threshold for our radio-loud versus radio-quiet classification. 
    }
    \label{fig:radio_R_vs_P}
\end{figure}

\begin{figure}
    \centering
	\includegraphics[width=\columnwidth]{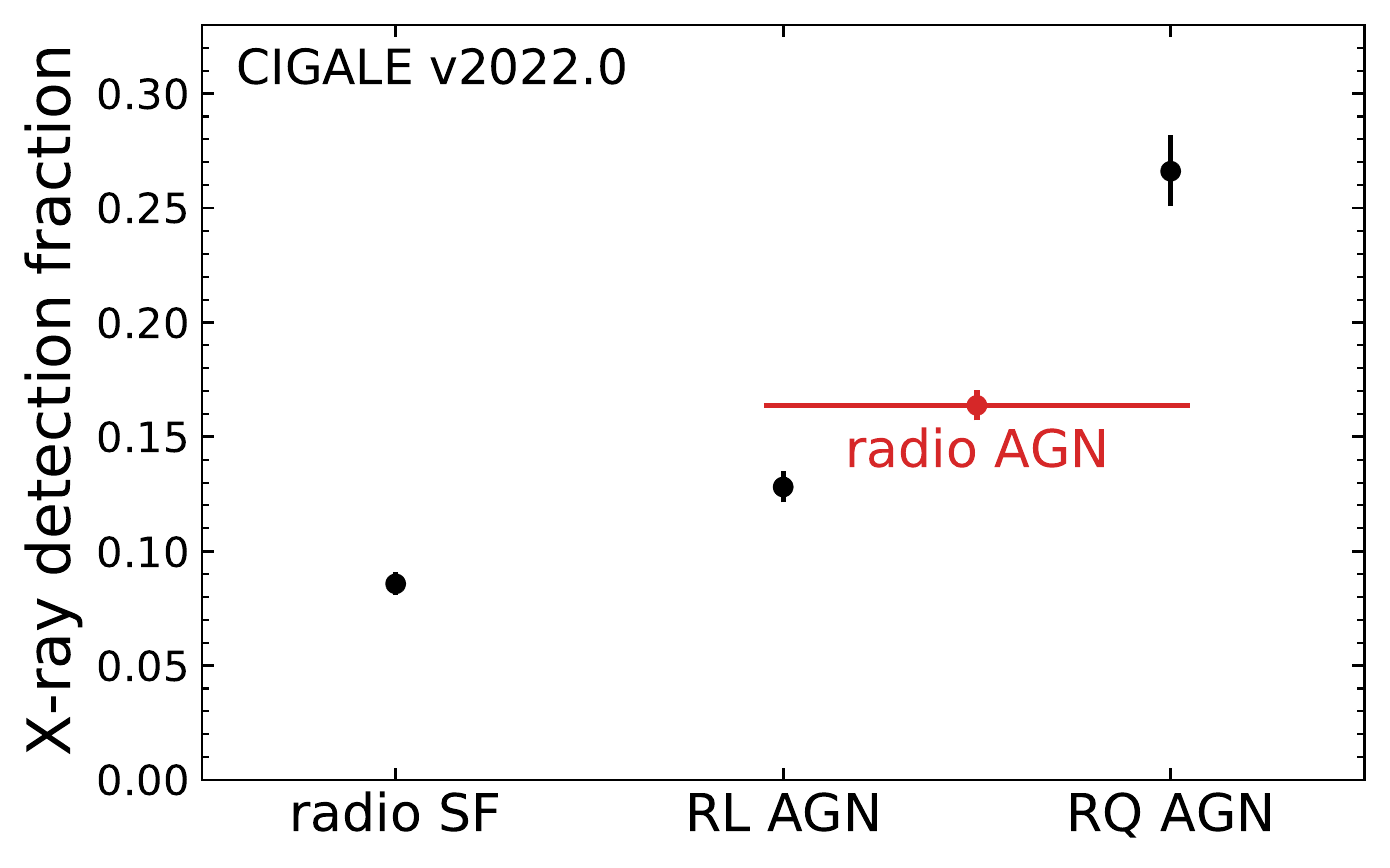}
    \caption{\xray\ detection fraction for different radio source types (as labeled).
    The red data point represents the radio AGN sample ($\pagn/\delta \pagn>2$), including both RL and RQ AGNs. 
    \fst{The error bars represent binomial uncertainties.}
    The radio AGNs have a higher \xray\ detection fraction than the SF galaxies, suggesting a link between AGN radio and \xray\ emission. 
    }
    \label{fig:radio_xray_frac}
\end{figure}

Using \cigv, we re-fit the photometric data of the COSMOS radio sources (\S\ref{sec:radio_mot}).
We set $R_{\rm AGN}$ to a wide logarithmic-spaced grid from 0.01 to 10000 (see Table~\ref{tab:par_radio}) based on the observations of quasars \citep[e.g.,][]{zhu20}.
\fst{We fix the radio slope in Eq.~\ref{eq:alpha_agn} at the default value of $\alpha_{\rm AGN}=0.7$, as
most (73\%) of our sources only have one radio band (3 GHz) available.}

Unlike the results from \xcig, the model radio fluxes agree well with the observed fluxes (see \fst{ Figs.~\ref{fig:radio_obs_vs_mod} \& \ref{fig:radio_1p4G_obs_vs_mod}}). 
The offsets between model and observed radio fluxes are \fst{mostly (99.91\% for 3~GHz and 98.6\% for 1.4~GHz)} within 0.5~dex. 
Therefore, we conclude that our implementation of the AGN radio component is indeed useful in explaining the observed radio flux, \fst{although the good fit of radio data is not surprising given only one (or two) radio band(s) is available for each source in our sample.
Interestingly, the model 1.4~GHz fluxes tend to be slightly lower than the observed ones (median offset $=0.08$~dex), suggesting that the typical AGN radio SED in the COSMOS sample is steeper than our assumed $\alpha_{\rm AGN}=0.7$ (Table~\ref{tab:par_radio}).
However, this systematic offset might also be a selection effect, as the relatively shallow 1.4~GHz data may miss AGNs with flatter radio SEDs and thereby lower 1.4~GHz fluxes. 
}

Fig.~\ref{fig:radio_sed} compares example SED fits from \xcig\ and \cigv. 
The observed radio fluxes are dominated by the AGN component from the \cigv\ fit. 
The \xcig\ fit is not able to explain the radio fluxes due to the lack of AGN radio emission. 
Compared to the \cigv\ fit, the \xcig\ fit has a stronger galaxy IR component. 
This is because \xcig\ only has galaxy radio emission which is related to galaxy IR emission through the radio-IR correlation (\S\ref{sec:radio_mot}). 
The high observed radio flux forcibly elevates not only galaxy radio emission but also its IR emission as a consequence.
In the \cigv\ fit, the radio flux is mostly explained by the AGN component, and thus the strong requirement of a galaxy component is relaxed. 

{\sc cigale v2022.0}\ can calculate AGN rest-frame 1.4~GHz luminosity ($\pagn$; e.g., \citealt{padovani16}) as a measure of AGN radio strength. 
In this work, we consider the sources with $\pagn/\delta \pagn>2$ (where $\delta \pagn$ is the $\pagn$ uncertainty from \cigv) as radio AGNs and the rest as radio SF galaxies. 
This definition guarantees that the AGN radio component is statistically significant ($>2\sigma$) for the classified radio AGNs.
We note that our definition of AGN/SF is based on the radio-band decomposition, because our focus here is radio emission. 
For example, if a source has AGN features at other wavelengths (e.g., \xray; see below) but its AGN radio emission is insignificant, it will be classified as a radio SF galaxy here. 
There are a total of 3221 radio AGNs, 50\% of the sample. 
This high fraction indicates that radio AGNs are common among the sources selected by deep radio surveys. 
The radio-AGN fraction depends on radio fluxes. 
The fractions are 47\% and 85\% for sources with 3~GHz fluxes below and above 0.2~mJy, respectively. 
This significant radio-flux dependence is also found by \citet{smolcic17b}, who used empirical criteria to classify AGNs and SF galaxies. 

Fig.~\ref{fig:radio_R_vs_P} displays $R_{\rm AGN}$ versus $\pagn$ and their distributions for our radio AGNs ($\pagn/\delta \pagn>2$). 
The red dashed line marks the conventional threshold (i.e., $R_{\rm AGN}=10$; \citealt{kellermann89}) for radio-quiet (RQ) versus radio-loud (RL) AGN classifications. 
The numbers of RQ and RL AGNs are 823 (26\%) and 2398 (74\%), respectively. 
We remind the reader that this RQ/RL classification demonstrates an advantage of \cigv, which simultaneously models multiwavelength data in a consistent way (\S\ref{sec:radio_code}). 
Such a task is challenging for empirical approaches, because the AGN UV/optical emission is often heavily obscured and not directly observable (e.g., Fig.~\ref{fig:radio_sed}).

X-ray emission is a good tracer of the BH-accretion process \citep[e.g.,][]{brandt15, brandt21}. 
It is intriguing to investigate the \xray\ emission of our classified radio types. 
We adopt the \chandra\ COSMOS-Legacy survey \citep{civano16, marchesi16}.   
A total of 801 of the radio-selected sources in our radio sample are detected in \xray. 
Fig.~\ref{fig:radio_xray_frac} displays the fractions of \xray\ detected sources among different radio types. 
The error bars represent binomial uncertainties calculated using {\sc astropy.stats.binom\_conf\_interval}.
The uncertainties are negligible compared to the differences across different radio types, thanks to our relatively large sample sizes.

The \xray\ fraction of the radio AGNs is 1.9~times higher than that of the radio SF. 
This result indicates that there is a positive link between AGN radio and \xray\ emission, \fst{broadly consistent with the findings in the literature \citep[e.g.,][]{merloni03, laor08}}.
Among the radio AGNs, the RQ population has a higher \xray\ detected fraction than the RL population (Fig.~\ref{fig:radio_xray_frac}). 
This is expected, because RQ should have higher $\luvr$ than RL at a given $\pagn$ (see Eq.~\ref{eq:R_agn}), 
and $\luvr$ is strongly correlated with AGN $\lx$ due to the $\ox$-$\luvr$ relation \citep[e.g.,][]{steffen06,just07}.
However, the \xray\ fraction of the RL AGNs is still significantly higher than that of the radio SF population (13\% vs.\ 9\%). 
Assuming that the radio emission in RL AGNs is mainly from jets (\S\ref{sec:agn_radio}), the elevated \xray\ fraction of the RL population suggests a connection between jets and \xray\ emission. 
This connection suggests that AGN jets could actively produce X-rays \citep[e.g.,][]{harris06}, or that there is a positive link between jets and the \xray\ emitting coronae \citep[e.g.,][]{zhu20}.

We note that, since we do not include the \xray\ data in our \xcig\ run (\S\ref{sec:radio_sample}), the \xray\ detection fraction is independent of our SF, RQ, and RL classifications. 
Therefore, the \xray\ fraction dependence on the radio type should be intrinsic, not a bias due to our SED-fitting procedure.

We set the default $\alpha_{\rm AGN}=0.7$ (see \S\ref{sec:radio_code}) and $R_{\rm AGN}=10$ (i.e., the boundary between RL and RQ AGNs). 
For $\alpha_{\rm AGN}$, when there are multi-frequency radio data spanning a large wavelength range, the user can adopt multiple $\alpha_{\rm AGN}$ values to better model the observed radio fluxes. 
When only one or two radio bands within a narrow wavelength range (like our case) are available, the user can just keep the default $\alpha_{\rm AGN}$ to save memory and reduce computational time. 
\fst{For the parameter of $R_{\rm AGN}$, we recommend the user to adopt multiple values based on our fits (e.g., Fig.~\ref{fig:radio_R_vs_P}).}  
The user can narrow the range of $R_{\rm AGN}$ in some cases. 
For example, if the sources have spatially extended radio structures (strong evidence for radio-loud AGNs), then $R_{\rm AGN}$ can be set to $>10$ values.



\section{Miscellaneous Updates}
\label{sec:misc}
\begin{figure}
    \centering
	\includegraphics[width=\columnwidth]{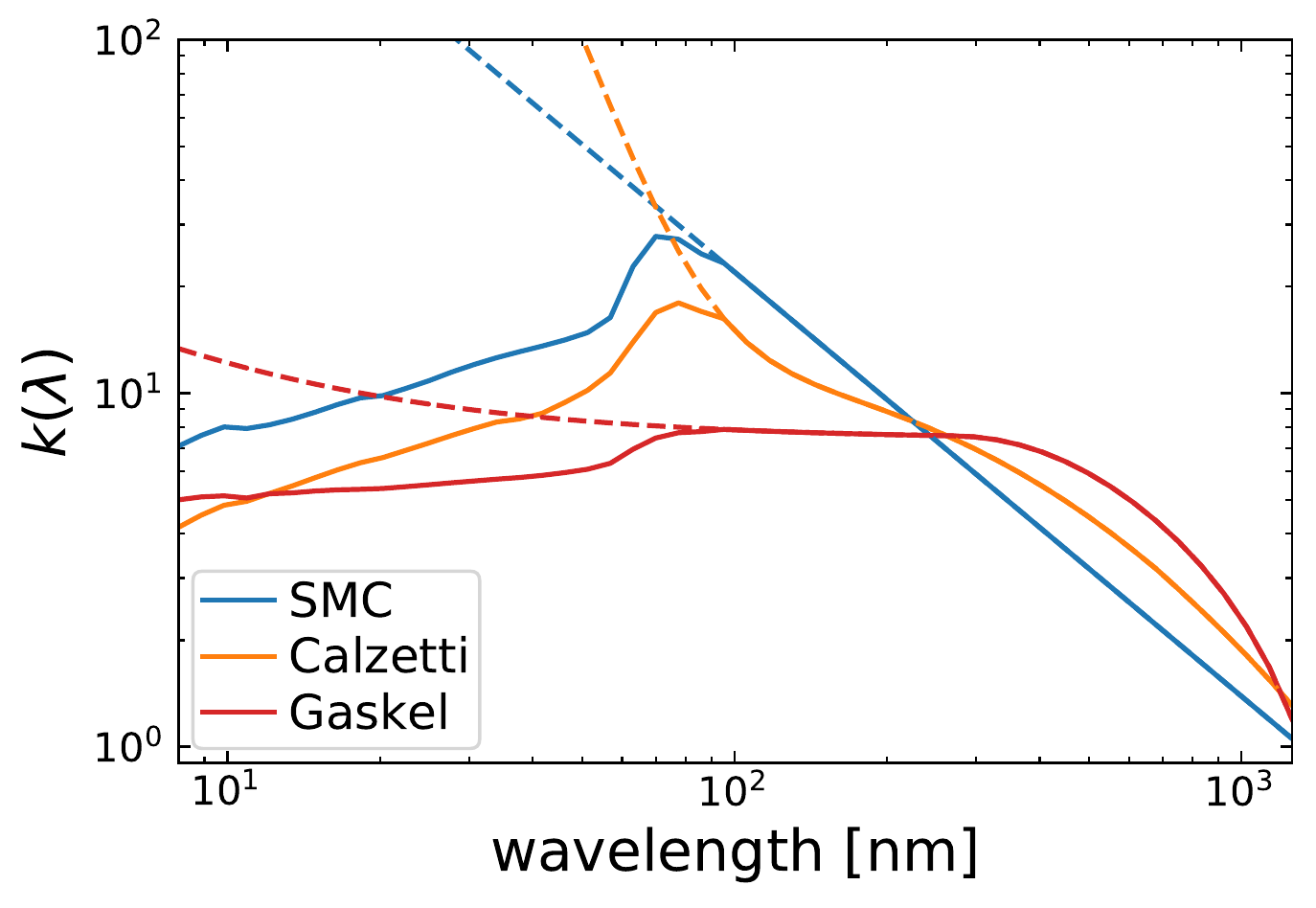}
	\includegraphics[width=\columnwidth]{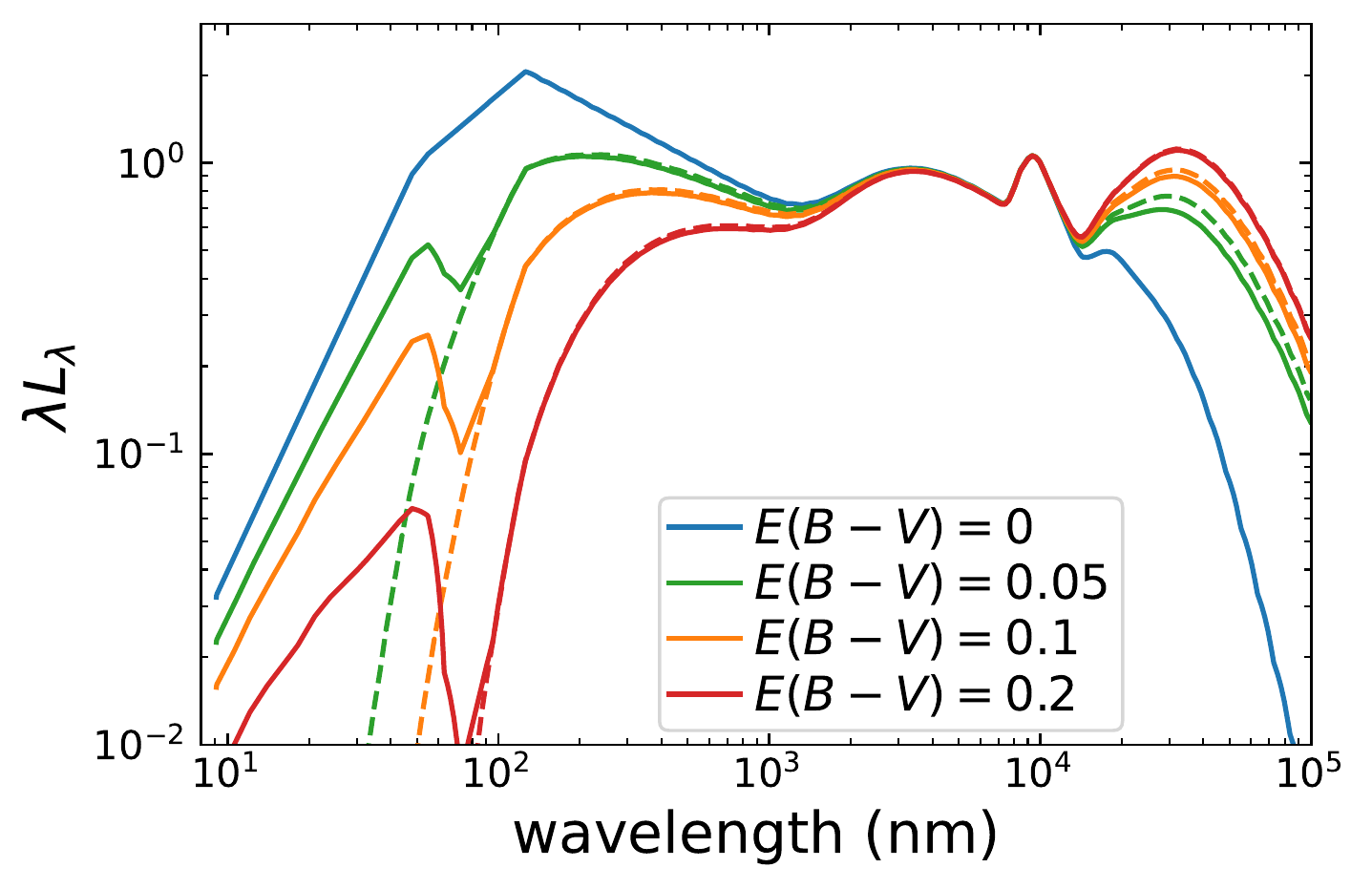}
    \caption{\textit{Top}: The extinction curves in \cigv\ (solid) and \xcig\ (dashed). 
    Different colors indicate different extinction laws as labeled. 
    Below 100~nm the \cigv\ curves have been obtained calculated based on the optical properties of the realistic dust mixtures, while the \xcig\ curves are analytical extrapolation.
    \textit{Bottom}: The type~1 AGN SED models with different polar-dust $E(B-V)$.
    The models are normalized at 10~$\mu$m.
    The solid and dashed curves are based on the \cigv\ and \xcig\ SMC extinction curves, respectively.
    The polar-dust re-emission ($\sim 30\ \mu$m) is lower (bottom panel) when using the \cigv\ extinction, because the \cigv\ extinction is weaker than the \xcig\ one at $<10$~nm and both codes follow energy conservation. 
    }
    \label{fig:klam}
\end{figure}

Besides the major changes of the code detailed in previous sections, we also implement several minor updates as below. 
\begin{itemize}
    \item In \xcig, the $\ox$ parameter (\xray\ module) is internally set to $-1.9$, $-1.8$, $-1.7$, ..., $-1.1$, and the user cannot control it.
    In \cigv, we change $\ox$ to an explicit parameter that is set by the user, although the default values are still $-1.9$, $-1.8$, ..., $-1.1$. 
    
    This change allows the user to run \xcig\ more effectively when using the \xray\ module. 
    For example, if the sample \fst{consists of} luminous quasars which typically have more negative values of $\ox$ \citep[e.g.,][]{just07}, then the user can set $\ox$ as, e.g., $-1.9$, $-1.8$, $-1.7$, and $-1.6$.
    This setting will reduce the number of models by a factor of 2.25, significantly boosting the efficiency. 
    
    This update of $\ox$ also allows the investigations of rare AGNs that have extreme $\ox$ values. 
    For example, the class of \xray\ weak quasars can have $\ox < -1.9$ \citep[e.g.,][]{pu20}, beyond the fixed $\ox$ parameter grid in \xcig. 
    In \cigv, the user can adopt $\ox$ more negative than $-1.9$ to probe the \xray\ weak population. 
    
    \item In \xcig, the normalization of the AGN component is controlled by the parameter of AGN fraction, 
    defined as $\fracA=\frac{L_{\rm dust, AGN}}{L_{\rm dust, AGN}+L_{\rm dust, galaxy}}$, where $L_{\rm dust, AGN}$ and $L_{\rm dust, galaxy}$ are AGN and galaxy dust luminosity (integrated over all wavelengths), respectively. 
    In \cigv, we allow the user to change the definition wavelength (range) of $\fracA$ by another parameter, ``lambda\_fracAGN''.  
    Setting it to ``$\lambda_{\rm min}/\lambda_{\rm max}$'' (units: $\mu$m) means that $\fracA$ is defined as $\frac{L_{\rm AGN}}{L_{\rm AGN}+L_{\rm galaxy}}$, where $L_{\rm AGN}$ and $L_{\rm galaxy}$ are AGN and galaxy total luminosity (not only dust) integrated over the wavelength range from $\lambda_{\rm min}$ to $\lambda_{\rm max}$.
    If $\lambda_{\rm min}=\lambda_{\rm max}$, then the code will use the monochromatic luminosity at this wavelength. 
    If lambda\_fracAGN is set to ``0/0'' (the default values), then the code will still use the definition of $\fracA$ in \xcig. 
    This change allows the users to model AGN versus galaxy relative strength in their interested wavelengths. 
    
    \item {\sc x-cigale} allows three extinction laws for AGN polar dust, i.e., \citet[][nearby star-forming galaxies]{calzetti00}, \citet[][large dust grains]{gaskell04}, and \citet[][Small Magellanic Cloud, SMC]{prevot84}. These extinction laws extend to $\approx 100$~nm, below which \xcig\ adopts analytical extrapolations. These extrapolations lead to large, non-physical extinctions below $100$~nm. This has a direct impact in the models, the dust reprocessing too much radiation that is re-emitted in the infrared while dramatically steepening the slope of the accretion-disk emission. 
    In order to address this issue, the extinction curves were recalculated in the entire wavelength range of interest by the module within the {\sc skirt} radiative-transfer code \citep{Baes2011,Baes-Camps2015,Camps-Baes2015} based on the realistic dust mixtures and optical properties taken from the literature. The SMC dust mixture consists of populations of silicate and graphite dust grains. The grain size distribution is taken from \citet{Weingartner-Draine2001}: a power-law function with a curvature and an exponential cutoff.
    Instead of extrapolation, below 100~nm, Calzetti extinction curve was replaced by the one corresponding to the standard Galactic interstellar dust \citep{MRN1977}.
    The Gaskell dust mixture represents a modification of the \citet{MRN1977} consisting of silicate and graphite populations with power-law grain size distribution: abundance of graphite is lowered to $15\%$, power-law exponent is taken to be $-2.05$ and the maximum grain size is lowered to 0.2 \micron.
    We adopt these new extinction curves below 100~nm for \cigv.
    In Fig.~\ref{fig:klam} we \fst{display} the \xcig\ and \cigv\ extinction curves (top) as well as some example type~1 AGN models with different polar-dust $E(B-V)$ (bottom).
    
\end{itemize}

\section{Summary and Future Prospects}
\label{sec:sum}
In this work, we test \xcig\ on different AGN/galaxy samples and improve the code accordingly. 
We publicly release the new code as \cigv\ on \url{https://cigale.lam.fr}. 
Our main results are summarized below. 
\begin{itemize}
    \item The \xcig\ fits of COSMOS type~2 AGNs produce systematically negative $\Delta \ox$, indicating that the observed \xray\ fluxes are below the expectations from the isotropic AGN \xray\ model. 
    In \cigv, we allow the user to model AGN $\lx$ \fst{(intrinsic \xray\ luminosity)} as a 2nd-order polynomial of $\cos \theta$. 
    We test three different sets of polynomial coefficients, i.e., $(a_1, a_2)=$(0.5, 0), (1, 0), and (0.33, 0.67), and compare the results with that of the isotropic model. 
    We find that the fits from $(a_1, a_2)=$(0.5, 0) have the best quality in terms of both $\Delta \ox$ and $\Delta $AIC.
    This result indicates that AGN \xray\ emission is moderately anisotropic in general (see \S\ref{sec:xray_ani}).
    
    \item For the CDF-S normal galaxies, the model \xray\ fluxes from \xcig\ do not agree well with the observed fluxes for many sources, e.g., 21\% of the sources have offsets $>0.5$~dex. 
    These offsets reflect the intrinsic scatters of the $\lhmxb$-SFR and $\llmxb$-$\mstar$ scaling relations for individual galaxies due to, e.g., globular clusters and statistical fluctuations. 
    Therefore, in \cigv, we introduce two new free parameters, $\dethmxb$ and $\detlmxb$, which are the logarithmic deviations from the default $\lhmxb$-SFR and $\llmxb$-$\mstar$ scaling relations. 
    We set both parameters from $-0.5$ to 0.5 with a step of 0.1 and re-fit the sources with \cigv. 
    All of the resulting model fluxes agree with the observed fluxes within 0.5~dex. 
    The resulting $\dethmxb$ and $\detlmxb$ distributions both show a slightly positive trend, suggesting a systematic offset of the $\lhmxb$ and $\llmxb$ scaling relations or an \xray\ selection bias (see \S\ref{sec:xrb}). 
    
    \item A significant fraction (32\%) of SDSS quasars have $u-z$ colors bluer than the model limit ($u-z=0.5$) of \xcig. 
    We allow the user to adjust the UV/optical slope of the intrinsic disk model with a $\delta_{\rm AGN}$ parameter in \cigv.
    This change successfully models the observed blue quasar SEDs. 
    The fitted $\delta_{\rm AGN}$ has a negative median value ($-0.27$), suggesting that the typical intrinsic quasar SED might be bluer than the default \citet{schartmann05} model.
    However, the degeneracy between $\delta_{\rm AGN}$ and $E(B-V)_{\rm PD}$ might also contribute to this negative trend (see \S\ref{sec:flex_disk}). 
    
    \item {\sc x-cigale} only accounts for galaxy radio emission. 
    Its fits of COSMOS radio sources fail to account for the observed radio 3~GHz fluxes in many cases, e.g., 28\% of the sources have model fluxes more than 0.5 dex below the observed ones. 
    Therefore, in \cigv, we add an AGN radio \fst{power-law} component, parameterized by AGN loudness ($R_{\rm AGN}$) and power-law slope ($\alpha_{\rm AGN}$).
    With this AGN component, \fst{the model agrees with the observed 3~GHz (and 1.4~GHz when available) data point within 0.5~dex for most sources}. 
    From the fits of \cigv, we find that about half of the radio sources have a significant radio AGN component (as defined by $\pagn/\delta \pagn>2$), and we classify the rest as radio SF galaxies.  
    \fst{This result suggests that AGN activity is common 
    among sources selected by deep radio surveys.}

    \item We also implement several miscellaneous updates in \cigv. 
    We allow the user to set the AGN $\ox$ grid instead of fixing it.
    We introduce a new free parameter ``lambda\_fracAGN'' which sets the wavelength range for $\fracA$ definition. 
    We improve the AGN polar-dust extinction curves at $\lambda \lesssim 100~nm$ based on realistic dust mixtures and optical properties taken from the literature. 
    These updates make \cigv\ more flexible and physical (see \S\ref{sec:misc}). 
 
\end{itemize}

Multiwavelength deep and/or wide surveys have become increasingly popular in extragalactic research. 
\cigv\ serves as a reliable and efficient tool to physically interpret the multiwavelength survey data from radio to \xray\ wavelengths. 
Its open-source nature and module-based structure \citep{boquien19} will also benefit the community. 
In our experience, most of user-specific needs can be satisfied by the original code or with slight/straightforward modifications. 
Future works can apply \cigv\ to current/ongoing surveys e.g., VLASS \citep{lacy20}, eRASS \citep{predehl21}, and LoTSS \citep{shimwell17} as well as future surveys from, e.g., \jwst, \textit{Xuntian}, \athena, and SKA. 



\section*{Acknowledgments}
We thank the referee for helpful feedback that improved this work. 
We thank Mark Dickinson and Fan Zou for helpful discussions. 
GY and CP acknowledge support from the George P.\ and Cynthia Woods Mitchell Institute for Fundamental Physics and Astronomy at Texas A\&M University, from the National Science Foundation through grants AST-1614668 and AST-2009442, and from the NASA/ESA/CSA James Webb Space Telescope through the Space Telescope Science Institute, which is operated by the Association of Universities for Research in Astronomy, Incorporated, under NASA contract NAS5-03127.
MB acknowledges support by the ANID BASAL project FB21000 and the FONDECYT regular grants 1170618 and 1211000.
WNB acknowledges the support from NASA grant 80NSSC19K0961.
KM has been supported by the Polish National Science Centre (UMO-2018/30/E/ST9/00082). 
GM acknowledges support by the Agencia Estatal de Investigaci\'{o}n, Unidad de Excelencia Mar\'{i}a de Maeztu, ref. MDM-2017-0765. 
MS acknowledges support by the Ministry of Education, Science and Technological Development of the Republic of Serbia through the contract no. 451-03-9/2021-14/200002 and by the Science Fund of the Republic of Serbia, PROMIS 6060916, BOWIE.
The authors acknowledge the Texas A\&M High Performance Research Computing Resources (HPRC, \url{http://hprc.tamu.edu}) that contributed to the research reported here.

\software{
{\sc astropy} \citep[v4.2][]{astropy},
{\sc x-cigale} \citep{boquien19, yang20}.
}


\bibliography{all}{}
\bibliographystyle{aasjournal}



\end{CJK*}
\end{document}